\documentclass[a4paper,fleqn,usenatbib]{mnras}

\usepackage[T1]{fontenc}
\usepackage{ae,aecompl}

\usepackage{graphicx}	
\usepackage{amsmath}	
\usepackage{amssymb}	
\usepackage{import}
\usepackage{lmodern}
\RequirePackage{fix-cm}
\usepackage[english,french]{babel}

\usepackage{graphicx}	
\usepackage{amsmath}	
\usepackage{amssymb}	
\usepackage{grffile}
\usepackage{numprint}
\usepackage[squaren,Gray]{SIunits}

\newcommand{\ex}{\mathbf{e}_{\rm x}}
\newcommand{\ey}{\mathbf{e}_{\rm y}}
\newcommand{\ez}{\mathbf{e}_{\rm z}}

\newcommand{\rlight}{r_{\rm L}}

\newcommand{\LL}{Landau-Lifshits }
\newcommand{\CP}{circular polarization }
\newcommand{\LP}{linear polarization }

\DeclareRobustCommand{\rchi}{{\mathpalette\irchi\relax}}
\newcommand{\irchi}[2]{\raisebox{\depth}{$#1\chi$}} 

\title[Particle acceleration and radiation reaction]{Particle acceleration and radiation reaction in strong spherical electromagnetic waves}
\author[J. P\'etri]{
J. P\'etri\thanks{E-mail: jerome.petri@astro.unistra.fr}
\\
Universit\'e de Strasbourg, CNRS, Observatoire astronomique de Strasbourg, UMR 7550, F-67000 Strasbourg, France.
}

\date{Accepted XXX. Received YYY; in original form ZZZ}

\pubyear{2021}

\begin{document}
\label{firstpage}
\pagerange{\pageref{firstpage}--\pageref{lastpage}}
\maketitle

\begin{abstract}
Strongly magnetized and fast rotating neutron stars are known to be efficient particle accelerators within their magnetosphere and wind. They are suspected to accelerate leptons, protons and maybe ions to extreme relativistic regimes where the radiation reaction significantly feeds back to their motion. In the vicinity of neutron stars, magnetic field strengths are close to the critical value of $B_{\rm c} \sim \numprint{4.4e9}$~T and particle Lorentz factors of the order $\gamma \sim 10^9$ are expected. In this paper, we investigate the acceleration and radiation reaction feedback in the pulsar wind zone where a large amplitude low frequency electromagnetic wave is launched starting from the light-cylinder. We design a semi-analytical code solving exactly the particle equation of motion including radiation reaction in the Landau-Lifshits approximation for a null-like electromagnetic wave of arbitrary strength parameter and elliptical polarization. Under conventional pulsar conditions, asymptotic Lorentz factor as high as $\numprint{e8}-\numprint{e9}$ are reached at large distances from the neutron star. However, we demonstrate that in the wind zone, within the spherical wave approximation, radiation reaction feedback remains negligible.
\end{abstract}

\begin{keywords}
	magnetic fields - methods: analytical - stars: neutron - stars: rotation - pulsars: general
\end{keywords}



\section{Introduction}

Strong magnetic fields dragged by fast rotation induce huge electric fields able to accelerate charged particles to ultra-relativistic speeds. Such conditions are met around strongly magnetized and fast spinning neutron stars known as pulsars and magnetars. These compacts astrophysical objects are indeed suspected to fill the interstellar and intergalactic medium with the most energetic particles in the universe and maybe also to produce part of the ultra high energy cosmic rays. These ideas where for instance explored by \cite{gunn_acceleration_1969} by using a vacuum wave and then improved by \cite{kegel_acceleration_1971} assuming a refractive index different from vacuum. These ideas were also revisited by \cite{thielheim_cosmic-ray_1990}. It is still unclear where and how efficient such acceleration mechanisms are around neutron stars. However, three main regions have been identified: the inner magnetosphere, that is the corotating quasi-static zone \citep{goldreich_pulsar_1969}, the wind zone \citep{coroniti_magnetically_1990, michel_magnetic_1994} where a low frequency large amplitude electromagnetic wave is launched and the termination shock of the pulsar wind \citep{petri_magnetic_2007}. Alternatively, magnetized relativistic outflows can also produce high energy particles via the Fermi process, diffusive shock acceleration, shock drift acceleration or magnetic reconnection, see for instance the review by \cite{matthews_particle_2020}.

In this work, we focus on particle acceleration by large amplitude electromagnetic waves. Relativistic acceleration of charged particles with mass~$m$ and charge~$q$ in a plane electromagnetic wave reveals efficient when the strength parameter defined by~$a = \omega_{\rm B}/\omega$ becomes much larger than one. Here $\omega_{\rm B}$ is the particle cyclotron frequency and $\omega$ the wave frequency. 
The strength parameter~$a$ gives a first guess for the energy gained by a particle starting from rest when accelerating in the electromagnetic field during one period of the wave. To orders of magnitude, the particle momentum divided by its mass is $\gamma\,\beta \approx a$ where $\beta$ is the normalized velocity with respect to the speed of light and $\gamma$ the associated Lorentz factor. As typical values for this strength parameter~$a$, we remember that for visible light, taking a wavelength of $\lambda=1~$\SIunits{\micro\meter} and a flux of $1$~\SIunits{\watt\per\meter\squared} corresponding to a magnetic field of \numprint{2e-6}~\SIunits{\tesla}, it amounts to
\begin{equation}
a \approx \numprint{e-10} \ll 1.
\end{equation}
Such optical waves are therefore unable to accelerates particles to even mildly relativistic speeds.
For current state technology with laser power of $\numprint{e24}$~\SIunits{\watt\per\meter\squared}, it becomes significantly larger than one and up to values about
\begin{equation}
a \approx \numprint{e3} .
\end{equation}
Mildly relativistic regimes are reachable by current state-of-the-art technology.
It is even expected to be soon possible to study radiation reaction effects during electron acceleration phases and to test the Lorentz-Abraham-Dirac (LAD) prescription for the charged particle equation of motion subject to radiation reaction. The correction term brought to the Lorentz force introduced by \cite{abraham_prinzipien_1902, abraham_zur_1904} and reinvestigated by \cite{lorentz_theory_1916} was eventually formulated in the relativistic regime by \cite{dirac_classical_1938}. This so-called LAD equation is still awaiting for experimental support and verification. It is known to be subject to run-away solutions that must be discarded. There exist an extensive literature on this topic, see for instance \cite{rohrlich_classical_2007} for a summary or also alternative radiation reaction contributions like the one deduced by \cite{eliezer_classical_1948}. In the astrophysical context of strongly magnetised rotating neutron stars, for instance for the archetypal Crab pulsar the strength parameter can reach extremely large values as high as
\begin{equation}
a = \numprint{e18} \gg 1
\end{equation}
at the stellar surface and somewhat lower at the light-cylinder ($\rlight=c/\omega$), about \numprint{e9} but still extremely high. Pulsars are therefore excellent candidates to push particles to ultra-relativistic energies by producing an electromagnetic kick on a very short time scale. The strength parameter at the light-cylinder, where the wave emerges has decreased by several orders of magnitude but remains significantly larger than one depending on the period~$P$ and its derivative~$\dot{P}$
\begin{equation}\label{key}
a_{\rm L} \approx \frac{q\,B}{m\,\omega} \, \left( \frac{R}{\rlight}\right)^3 \approx \numprint{4.5e7} \, \left( \frac{\dot{P}}{\numprint{e-15}} \right)^{1/2} \, \left( \frac{P}{\numprint{1}~\SIunits{\second}} \right)^{-3/2}
\end{equation}
thus still very large, $a_{\rm L} \gg 1$. In the present investigation, we study particle motion starting from this remote region up to very large distances, $r\gg \rlight$ where a plane wave is a very good approximation. Quantitative accurate results will be derived thanks to exact analytical solutions of the particle 4-velocity in a plane electromagnetic wave in the so-called \LL approximation derived by \cite{piazza_exact_2008} and retrieved by \cite{hadad_effects_2010} in a different form. However, the \LL approximation being a first order expansion of the Lorentz-Abraham-Dirac equation, it is not the only possible choice, see for instance the review by \cite{hammond_relativistic_2010} and also the exposition of alternative theories by \cite{burton_aspects_2014}.

\cite{gunn_acceleration_1969} were among the first to recognize the potential of pulsars to produce high energy cosmic rays. Following simple arguments, they found an estimate for the maximum energy that was refined by \cite{ostriker_nature_1969}. 
\cite{laue_acceleration_1986} studied the acceleration of protons and electrons for an perpendicular rotator in the \LL approximation, showing their orbit and Lorentz factor evolution with time. They also give detailed maps of maximum energy depending on the initial longitude of the particle. See also \cite{leinemann_acceleration_1988} for similar ideas.
\cite{kegel_radiation_1995} studied acceleration and radiation of charged particles in strong electromagnetic waves using exact analytical solutions for linearly and circularly polarized waves. They also looked for cold plasma effects. Strongly magnetized rotating neutron stars are believed to be efficient cosmic ray accelerators \citep{thielheim_particle_1991} but a clear picture of what kind of particles and to which energies they can be accelerated is still lacking. \cite{thielheim_particle_1993} performed a careful analysis of particle acceleration in a spherical wave field produced by a rotating dipole. This work was continued by \cite{thielheim_plasma_1994} who also computed some plasma configurations. In the same vain, \cite{michel_electrodynamics_1999} studied particle motion in a plane wave and in the Deutsch field. \cite{tolan_charged_1992} presented approximated analytical solutions for a test particle evolving in a rotating magnetic dipole and around a pulsar. He showed that radiation reaction and gravitation are negligible compared to the geometrical effect of a decaying spherical wave in the pulsar wave zone.

The propagation of strong electromagnetic waves in dense plasmas, being uniform or showing gradients, was performed back to the 70s by \cite{max_strong_1971} where conditions for the transmission of a plane wave are given. The radiation damping of a strong linearly polarized wave launched by a pulsar and due to electron-positron pair synchro-Compton radiation was explored by \cite{asseo_synchro-compton_1978}. They showed that for conditions prevailing in the Crab pulsar, the wave fades away within only several wavelengths.

All these works focused on the large scale acceleration. On the opposite side, \cite{ferrari_acceleration_1974} computed particle acceleration and radiation in the near field region, very close to the neutron star surface, showing significant radiation reaction braking in this near zone. However, as the strength parameter~$a$ decreases due to the dipolar nature of the magnetic field, acceleration and radiation reaction become less and less effective.

Earlier works already worried about the effect of radiation reaction on particle acceleration. For instance  \cite{heintzmann_acceleration_1972} studied particle acceleration and radiation reaction in plane and spherical waves, see also \cite{grewing_acceleration_1975} for radiation effects in pulsar fields. \cite{grewing_acceleration_1973} then showed that the presence of a longitudinal magnetic field significantly reduces the maximum Lorentz factor of the accelerated particles. Surprisingly, radiation reaction is able to increase the asymptotic Lorentz factor of the charged particle when interacting for instance with an intense laser pulse as shown by \cite{fradkin_radiation_1979}. Synchrotron radiation spectra are also modified because of the decaying orbit of electrons in an uniform magnetic field as shown by \cite{nelson_synchrotron_1991}.

Radiation reaction is usually treated as a perturbation of the Lorentz force and called the \LL approximation. Therefore \cite{finkbeiner_applicability_1990} checked the validity of this approximation in pulsar vacuum fields which requires a classical description of the emitting particles, the radiation field and the smallness of the radiation reaction force compared to the Lorentz force in the particle instantaneous rest frame. This justifies the approach of integrating the particle equation of motion in the \LL limit for highly relativistic particles as performed by \cite{finkbeiner_effects_1989} starting from the neutron star surface. In strong electromagnetic fields, the quantum nature of the particles also emerges, leading to an additional equation for the evolution of the particle spin as implemented numerically by \cite{li_accurately_2020}.

Finding accurate and exact analytical solutions to the particle equation of motion is crucial in ultra-strong electromagnetic fields as shown by \cite{petri_relativistic_2020} in the Lorentz force limit. Some applications to neutron stars have been explored by \cite{tomczak_particle_2020}. Different approaches exist to tackle the problem of finding exact and efficient implementations of the Lorentz force equation, see for instance \cite{gordon_pushing_2017} and \cite{gordon_special_2021} who also discuss the possibility to add radiation reaction.

In this paper we study particle acceleration and radiation reaction in the wind zone, approximating the field locally by a plane wave with decreasing amplitude in order to mimic a spherical wave. Our integration of the test particle equation of motion relies on exact analytical solutions of the \LL equation for either time-dependent elliptically polarized plane waves or constant null like electromagnetic fields. These solutions are recalled in Sec.~\ref{sec:Solutions} and serve as a building block for our algorithm. As a first step towards a more general algorithm able to integrate semi-analytically any field configuration, we also try an algorithm based on locally constant electromagnetic field solutions. Both numerical schemes are then tested in plane polarized waves for the Lorentz force in Sec.~\ref{sec:Tests} and with radiation reaction in the \LL limit in Sec.~\ref{sec:Tests2}, showing the good agreement of the locally constant approximation with the analytical solution. We then discuss our new results about acceleration efficiency and final Lorentz factor of particles in a spherical waves in Sec.~\ref{sec:Results} including radiation reaction. The limitation of our present study focusing on null-like electromagnetic fields is discussed in Sec.~\ref{sec:Discussion}. Conclusions are drawn in Sec.~\ref{sec:Conclusions}.

\section{Exact solutions}\label{sec:Solutions}

Our aim is to solve the particle acceleration and radiation damping problem by time dependent numerical simulations, sticking as close as possible to known exact analytical solutions. We start from the linearised Lorentz-Abraham-Dirac equation leading to first order to the \LL prescription \citep{landau_physique_1989} such that
\begin{multline}\label{eq:LL}
\frac{du^i}{d\tau} = \frac{q}{m} \, F^{ik} \, u_k + \frac{q \, \tau_0}{m} \, \partial_\ell F^{ik} \, u_k \, u^\ell + \\
 \frac{q^2 \, \tau_0}{m^2} \, \left( F^{ik} \, F_{k\ell} \, u^\ell + ( F^{\ell m} \, u_m ) \, ( F_{\ell k} \, u^k ) \, \frac{u^i}{c^2} \right) .
\end{multline}
$q$ and $m$ are the particle charge and rest mass, $u^i$ its 4-velocity, $\tau$ its proper time, $F^{ik}$ the electromagnetic or Faraday tensor, $c$ the speed of light and $\tau_0$ the light crossing time across the electron classical radius~$r_e$ (within a factor unity)
\begin{equation}\label{eq:tau_0}
\tau_0 = \frac{q^2}{6\,\pi\,\varepsilon_0\,m\,c^3} = \frac{2}{3} \, \frac{r_e}{c} \approx \numprint{6.26e-24}~\SIunits{\second}.
\end{equation}
Fortunately, there exist some exact analytical solutions to this equation \eqref{eq:LL} in either a constant electromagnetic field or for an elliptically polarized plane wave depending on the two electromagnetic invariants
\begin{subequations}
	\label{eq:Invariants}
\begin{align}
I_1 & = E^2/c^2 - B^2 \\
I_2 & = \mathbf{E}\cdot \mathbf{B} /c
\end{align}
\end{subequations} 
where $\mathbf{E}$ and $\mathbf{B}$ are the electric and magnetic field respectively as measured by some inertial observer. 
The two important parameters defining the family of solutions are the strength parameter $a$ and the radiation reaction efficiency $\omega\,\tau_0$ according to the following definitions
\begin{subequations}
	\label{eq:Parametres}
	\begin{align}
	a & = \frac{\omega_{\rm B}}{\omega} \\
	b & = \omega\,\tau_0 \\
	\omega_{\rm B} & = \frac{q\,B}{m} = \frac{q\,E}{m\,c} .
	\end{align}
\end{subequations}
Introducing the weighted and normalized electromagnetic field tensor by~$\tilde{F}^{ik} = q \, F^{ik}/m\,\omega \propto a$ and a normalized time $\tilde{\tau} = \omega \, \tau$, the \LL equation~\eqref{eq:LL} is rewritten without dimensions as
\begin{multline}\label{eq:LLnormee}
\frac{d\tilde{u}^i}{d\tilde{\tau}} = \tilde{F}^{ik} \, \tilde{u}_k + b \, \tilde{\partial}_\ell \tilde{F}^{ik} \, \tilde{u}_k \, \tilde{u}^\ell + \\
b \, \left( \tilde{F}^{ik} \, \tilde{F}_{k\ell} \, \tilde{u}^\ell + ( \tilde{F}^{\ell m} \, \tilde{u}_m ) \, ( \tilde{F}_{\ell k} \, \tilde{u}^k ) \, \tilde{u}^i \right) .
\end{multline}
The ordering of the right-hand side terms are $\gamma\,a, \gamma^2\,a\,b, \gamma\,a^2\,b, \gamma^3\,a^2\,b$, $\gamma$ being the particle Lorentz factor. For our application to neutron stars, $a\gg1$ and $\gamma\gg1$, therefore the last term dominates the radiation reaction. This last term is $\gamma^2\,a\,b = \gamma^2\,\omega_{\rm B}\,\tau_0$ times the Lorentz force. Therefore radiation reaction force becomes dominant in the regime where $\gamma^2\,\omega_{\rm B}\,\tau_0 \gtrsim1$ thus
\begin{equation}\label{eq:RRdomine}
\gamma \gtrsim \numprint{9.5e5} \, \left( \frac{B}{\numprint{1}~\SIunits{\tesla}}\right)^{-1/2}   .
\end{equation}

For the remainder of this paper, we use a Cartesian coordinate system labelled by $(x,y,z)$ and the corresponding Cartesian basis $(\ex,\ey,\ez)$. Moreover, the plane wave propagates in the $x$ direction, with a frequency~$\omega$, has a wave-vector~$k$ and a polarization electric vector $\mathbf{E}$ in the $yOz$ plane. Thus by construction $B_x=E_x=0$, $\mathbf{E}\cdot \mathbf{B}=0$ and $E=c\,B$. As we remind in the next section, exact analytical solutions have been found for those waves.

\subsection{Elliptically polarized plane waves}

An exact analytical solution of the \LL equation has been given by \cite{piazza_exact_2008} and \cite{hadad_effects_2010}. For completeness, as our algorithm heavily relies on this solution, we recall it by adopting slightly different notations compared to \cite{hadad_effects_2010}.

Let us assume a plane electromagnetic wave in vacuum with wave number~$k$ and frequency (more properly called pulsation)~$\omega$ such that the vector potential~$A^i$ is given by the real part~$\Re$ of a complex potential $f(\xi)\,\varepsilon^i$
\begin{equation}\label{eq:Potentiel_Vecteur}
A^i = A_0 \, \Re[f(\xi)\,\varepsilon^i]
\end{equation}
with $A_0$ the potential amplitude. $f(\xi)$ is an arbitrary function of the phase given by $\xi = k^i\,x_i = \omega \, t - k\,x$, the four-position vector is $x^i=(c\,t,x,y,z)$, the four-wavenumber~$k^i=(\omega/c,k,0,0)$ and the space like polarization vector $\varepsilon^i$. The strength of the wave is given in term of the parameter $a_0$ defined by 
\begin{equation}\label{eq:strength_parameter}
a_0 = \frac{q\,A_0}{m\,c} .
\end{equation}
Note that it can be positive or negative depending on the particle charge. The solution for the 4-velocity is then expressed by introducing several functions as 
\begin{subequations}\label{eq:psi}
\begin{align}
\psi(\xi) & = \int_0^\xi \mathbf{\hat{A}}'(y) \cdot \mathbf{\hat{A}}'(y) \, dy \\
\label{eq:tau}
\tau(\xi) & = \frac{\xi}{\mathbf{k} \cdot \mathbf{u}_0} - \tau_0 \, a_0^2 \int_0^\xi \psi(y) \, dy \\
\label{eq:chi}
\rchi^i & = \int_0^\xi \hat{A}'^i(y) \, \psi(y) \, dy \\
\label{eq:ku0}
\mathbf{k} \cdot \mathbf{u}_0 & = \gamma \, \omega \, (1-\beta_{\rm x}^0)
\end{align}
\end{subequations}
where the prime in $\hat{A}'^i(y)$ denotes the derivative with respect to the argument $y$.
Therefore, the full solution for an arbitrary wave is
\begin{multline}
	\label{eq:vitesse_plane}
	\frac{u^i}{\mathbf{k} \cdot \mathbf{u}} = \frac{u_0^i}{\mathbf{k} \cdot \mathbf{u}_0} + \frac{a_0\,c}{\mathbf{k} \cdot \mathbf{u}_0} \, \left[ -(\hat{A} - \hat{A}_0)^i \right. \\ 
	\left. + \frac{k^i}{\mathbf{k} \cdot \mathbf{u}_0} \, \left( (\mathbf{\hat{A}} - \mathbf{\hat{A}}_0) \cdot \mathbf{u}_0 - a_0 \, c \, \frac{(\mathbf{\hat{A}} - \mathbf{\hat{A}}_0)^2}{2} \right) \right] \nonumber \\
	 + \tau_0 \, c \, \left[ -a_0 \,(\hat{A}' - \hat{A}'_0)^i + a_0^3 \, \rchi^i \right. \\
	 \left. + \frac{k^i}{\mathbf{k} \cdot \mathbf{u}_0} \left( a_0 \,(\mathbf{\hat{A}}' - \mathbf{\hat{A}}'_0) \cdot \mathbf{u}_0 - a_0^2 \, c \, \psi \right. \right. \\
	 \left. \left. - a_0^2 \, c \, (\mathbf{\hat{A}} - \mathbf{\hat{A}}_0) \cdot (\mathbf{\hat{A}}' - \mathbf{\hat{A}}'_0) - a_0^3 \, \mathbf{\rchi} \cdot \mathbf{u}_0 + a_0^4 \, c \, (\mathbf{\hat{A}} - \mathbf{\hat{A}}_0) \cdot \mathbf{\rchi} \right) \right] \nonumber \\
	+ \tau_0^2 \, c^2 \, k^i \,\left[ - a_0^2 \, \frac{(\mathbf{\hat{A}}' - \mathbf{\hat{A}}'_0)^2}{2} + a_0^4 \, (\mathbf{\hat{A}}' - \mathbf{\hat{A}}'_0) \cdot \mathbf{\rchi} + a_0^4 \, \frac{\psi^2}{2} - a_0^6 \, \frac{\rchi^2}{2} \right].
\end{multline}
For charged particles immersed in the neutron star electromagnetic field, outside the light-cylinder, the field converges to a elliptically polarized plane wave depending on the colatitude~$\theta$. It is linearly polarized at the equator $\theta=\upi/2$ and circularly polarized at the poles $\theta=0$ and $\theta=\upi$, showing any kind of elliptic polarization between the poles and the equator. Therefore, in order to keep the discussion as general as possible, we focus on elliptically polarized waves with a wave vector $k^i=(\omega/c,k,0,0)$ and being a linear superposition of a left-handed and right-handed elliptically polarized wave with characteristics
\begin{subequations}
	\begin{align}
	\varepsilon_{\rm c}^i & = (0,0,1,-i)/\sqrt{2} \\
	f_\pm(\xi) & = \sqrt{2} \, e^{\pm i\,(\xi-\xi_0)} \\
	A_\pm^i & = A_0 \, (0,0,\cos(\xi-\xi_0),\pm\sin(\xi-\xi_0)) \\
	E_\pm^i & = -\omega \, A_0 \, (0,0,-\sin(\xi-\xi_0),\pm\cos(\xi-\xi_0)) \\
	B_\pm^i & = k \, A_0 \, (0,0,\cos(\xi-\xi_0), \pm\sin(\xi-\xi_0)) = k\, A_\pm^i .
	\end{align}
\end{subequations}
$\xi_0$ is the initial phase of the wave and the sign $\pm$ refers to a left or right handed elliptical polarization.
The full vector potential is a sum of left and right-handed elliptically polarized waves such that
\begin{equation}\label{eq:Potentiel_Vecteur_Elliptique}
A^i = \alpha \, A_+^i + (1-\alpha) \, A_-^i = (0,0, \cos(\xi-\xi_0), (2\,\alpha-1)\,\sin(\xi-\xi_0))
\end{equation}
where $\alpha \in[0,1]$ with $\alpha=1/2$ for linearly and $\alpha=0$ or $\alpha=1$ for circularly polarized waves with opposite handedness.

For spatially varying waves like spherical waves emitted by rotating neutron stars, we need to integrate the 4-velocity $u^i$ to deduce the phase dependence of the 4-position. Noting that $u^i = dx^i/d\tau = (\mathbf{k}\cdot \mathbf{u}) \, dx^i/d\xi$ we find that 
\begin{equation}\label{eq:dxsdxi}
\frac{dx^i}{d\xi} = \frac{u^i}{\mathbf{k}\cdot \mathbf{u}}
\end{equation}
which can also be integrating analytically for elliptically polarized waves starting from the expression~(\ref{eq:vitesse_plane}).

In the special case of particle propagation in an elliptically polarized wave, deviation from the Lorentz force motion sets in when $a^2\,b\gtrsim1$ \citep{hadad_effects_2010}. We will indeed check in our simulations that this condition is required for significant radiation feedback.

\subsection{Constant fields}

Unfortunately, the most general electromagnetic field is not null-like (meaning $I_1=I_2=0$). The simplest generalization leading to a tractable analytical solution is for constant fields. Following the procedure described by \cite{heintzmann_exact_1973}, we introduce the electromagnetic tensor eigensystem solution such that eigenvalues $\lambda_i$ (possibly complex values) satisfy
\begin{equation}\label{eq:eigenvalues}
\lambda_i^2 = \frac{I_1 \pm \sqrt{I_1^2 +4\,I_2^2}}{2}
\end{equation}
If $I_2=0$, then at least two eigenvalues vanish. For a null-like field meaning $I_1 = I_2 = 0$, all eigenvalues vanish and solutions are given in the previous paragraph. If all eigenvalues~$\lambda_i$ are non zero then the associated eigenvectors $\psi_i$ are null-like, $\psi_i \cdot \psi_i = 0$ because of the antisymmetry of the electromagnetic tensor. Moreover they are explicitly given by
\begin{equation}\label{eq:eigenvectors}
(\psi_i)^k = \left(\frac{\lambda_i^2 \, E^2 + c^2\,I_2^2}{\lambda_i\,c}, \lambda_i^2 \, \mathbf{E} + c\,I_2\,\mathbf{B} + \lambda_i \, \mathbf{E} \wedge \mathbf{B} \right) .
\end{equation}
These eigenstates form a complete basis for the four dimensional velocity space.
The 4-velocity is then adequately projected onto this basis according to
\begin{equation}\label{eq:vitesse}
u^k(\tau) = \sum_{i=1}^{4} k(\tau) \, f_i \, (\psi_i)^k \, e^{\lambda_i\,\tau}.
\end{equation}
The $f_i$ are the components of the 4-velocity in the $\psi_i$ basis. They are deduced from the initial conditions $u^k(0) = \sum_{i=1}^{4} \, f_i \, \psi_i$ and the damping factor is
\begin{equation}\label{eq:damping_factor}
k(\tau) = \left( \sum_{i\neq j} f_i \, f_j \, (\psi_i \cdot \psi_j ) \, e^{(\lambda_i+\lambda_j)\,\tau} \right)^{-1/2} .
\end{equation}
In the absence of radiation reaction, this damping factor equals unity.
In the most general electromagnetic field configuration, both invariants are non-vanishing, there are four distinct eigenvalues, two real and two complex conjugated and the eigenvectors form a full basis for the velocity space justifying the above projection scheme. Such configurations are met around rotating magnetized neutron stars, from the magnetosphere, inside the light-cylinder (the static zone) through the light-cylinder, the transition zone down to the wave zone, outside the light cylinder. Therefore the solution (\ref{eq:vitesse}) is the most appropriate building block to construct numerical schemes integrating particle trajectories around strongly magnetized neutron stars. However, in the present study, we focus only on plane waves for which analytical solutions exist, allowing detailed quantitative comparisons and error estimates between the algorithm proposed here and the expected values.

If some eigenvalues vanish, expression~\eqref{eq:eigenvectors} cannot be applied straightforwardly. Special care is required in these limiting cases. Of particular interest is the case when~$I_1=I_2=0$. Then all the eigenvalues vanish, the field is null- (or light-like) and the eigensystem must be solved separately as shown in the previous paragraph. 

If~$I_1\neq0$ and $I_2=0$ two eigenvalues vanish and the other two are either real and opposite or purely complex and opposite depending on the sign of~$I_1$. If $I_1>0$, the electric field~$\mathbf{E}$ dominates, the solutions being real and given by~$\lambda_{1,2} = \pm \sqrt{I_1}$ representing a pure electric accelerating solution. If~$I_1<0$ the magnetic field~$\mathbf{B}$ dominates and $\lambda_{1,2} = \pm i\,\sqrt{-I_1}$ representing oscillatory solutions, a simple magnetic gyration in the appropriate electric drift frame. In this case of perpendicular electric and magnetic fields, two eigenvalues vanish and for a non vanishing magnetic field $\mathbf{B} \neq \mathbf{0}$ the eigenvectors of the two dimensional null space are 
\begin{equation}\label{eq:vecteur_propre}
\left( \frac{\omega}{c}, \frac{\omega}{c^2} \, \frac{\mathbf{E} \wedge \mathbf{B}}{B^2} + \mu \, \mathbf{B}\right)
\end{equation}
with $(\omega,\mu)\in\mathbb{R}^2$ two arbitrary and uncorrelated reals generating the two dimensional null space.

In the special case of a zero magnetic field the above expression~\eqref{eq:vecteur_propre} fails and a separate treatment is required. The eigenvalues are real and given by
\begin{equation}\label{eq:valeur_propre_E}
\lambda_i = (0, 0, -E/c, +E/c) .
\end{equation}
The associated eigenvectors are
\begin{subequations}\label{eq:vecteur_propre_E}
	\begin{align}
	(\psi_1)^k & = (0,\mathbf{k}_1) \\
	(\psi_2)^k & = (0,\mathbf{k}_2) \\
	(\psi_3)^k & = (-E,\mathbf{E}) \\
	(\psi_4)^k & = (+E,\mathbf{E})
	\end{align}
\end{subequations}
with $\mathbf{k}_1 \cdot \mathbf{E} = \mathbf{k}_2 \cdot \mathbf{E} = 0$ and $\mathbf{k}_1 \wedge \mathbf{k}_2 \propto \mathbf{E}$. The spatial vectors $\mathbf{k}_1$ and $\mathbf{k}_2$ span the spatial plan orthogonal to the electric field~$\mathbf{E}$.

In this paper, we are interested in null-like fields with $I_1=I_2=0$, corresponding to electromagnetic waves launched by a rotating magnetic dipole, as seen at large distances $r\gg\rlight$, well outside the light-cylinder. In this special case, all eigenvalues vanish $\lambda_i=0$ and the solution for elliptically polarized waves applies.

%

\subsection{Initial conditions}

The aforementioned formal solutions depend on several physical parameters that have been reduced to two normalized quantities, namely the strength~$a$ and the damping~$b$ parameters. In order to quantitatively find the exact solution, we need to impose the initial conditions given by the initial phase of the wave~$\xi_0$ and the initial velocity of the particle~$\mathbf{u}_0$ injected at phase~$\xi_0$. We will only consider initial velocities aligned with the wave propagation direction such that $u_0^i = \Gamma_0\,c\,(1,\beta_0,0,0)$ where $\Gamma_0$ is the initial Lorentz factor and $\beta_0$ the normalized spatial velocity.

Note that the strength parameter is a Lorentz invariant because the electromagnetic field~$\mathbf{F}$ (meaning $\mathbf{E}$ or $\mathbf{B}$) between two frames of relative velocity~$\bbeta_0$ transforms according to $\mathbf{F} = \mathcal{D} \, \mathbf{F}'$ with the Doppler factor $\mathcal{D} = 1/\Gamma_0 \, (1-\ex \cdot \bbeta_0)$. The frequency is Doppler shifted based on 
\begin{equation}\label{eq:Doppler_Frequence}
\omega = \mathcal{D} \, \omega'
\end{equation}
rendering the ratio $B/\omega$ constant and equal to $B'/\omega'$ for frame velocities aligned with the wave propagation direction. Therefore $a=a'$ is indeed a relativistic invariant. When the particle has an initial velocity such as $\mathbf{u}_0$ it suffices to transform to the instantaneous particle rest frame at the initial time, to compute the solution in this frame with a particle at rest and finally to transform position and velocity back to the observer frame. The Lorentz factor~$\gamma'$ in the rest frame is related to the Lorentz factor~$\gamma$ measured by the observer by
\begin{equation}\label{eq:gammap1}
\gamma' = \gamma \, \Gamma_0 \, (1-\bbeta \cdot \bbeta_0)
\end{equation}
which simplifies for ultra-relativistic particles and $\bbeta $ aligned with $ \bbeta_0$ to
\begin{equation}\label{eq:gammap2}
\gamma' \approx \gamma \, \sqrt{\frac{1-\beta_0}{1+\beta_0}}.
\end{equation}
The period as measured by the observer also suffers from the time dilation effect. The Doppler effect for the wave frequency combined with the Lorentz transform for the time interval between two periods 
\begin{equation}\label{eq:cdt}
c\,\Delta t = \gamma \, (c\,\Delta t' + \beta \, \Delta x')
\end{equation}
corresponding to $\xi=2\,\upi$ shows that the period is changed to 
\begin{equation}\label{eq:Periode_Circ}
\omega \, T_{\rm circ} = 2\,\upi\,\mathcal{D} \, \gamma \, [1 + (\beta+1) \, a^2 ]
\end{equation}
for a circularly polarized wave and to 
\begin{equation}\label{eq:Periode_Lin}
\omega \, T_{\rm lin} = 2 \, \upi \, \mathcal{D} \, \gamma \, [1 + \frac{3}{4}\,(\beta+1) \, a^2 ]
\end{equation}
for a linearly polarized wave (to be compared with particles starting at rest, see \cite{petri_relativistic_2020}). We will check this point of view in the numerical tests discussed in Sec.~\ref{sec:Tests}.

If the initial phase~$\xi_0$ of the wave at the particle injection point does not vanish, the acceleration process is not optimal in the sense that the highest Lorentz factor~$\gamma$ will be less than $\gamma_{\rm max} = 1+2\,a^2$. For instance for the motion without radiation reaction, the particle is insensitive to the initial phase of a circularly polarized wave but sensitive to an elliptically polarized wave, the worst case being a linearly polarized wave with $\alpha=1/2$. For those waves the maximum Lorentz factor is $\gamma_{\rm max}(\xi_0) = 1+\frac{1}{2}\,a^2\,(1+\|\cos\xi_0\|)^2$ therefore a factor 4 less for $\xi_0=\upi/2$ compared to the optimal case $\xi_0=0$ when $a\gg1$. This will also be checked in our subsequent tests.

\section{Plane wave tests without radiation reaction} \label{sec:Tests}

In this section, we perform some tests of the constant field approximation for plane waves and compare our results with the exact analytical solutions detailed in the previous section. We distinguish cases with particles initially at rest from cases with particles initially moving at relativistic speed catching up the wave or moving in opposite direction to the wave. We then close the test section by a discussion of the impact of the initial phase of the wave on the acceleration efficiency. The phase is indeed another important parameter controlling the maximum energy reached by the particle.

\subsection{Particle starting at rest}

Let us assume that particles are injected at rest in an electromagnetic wave with an initial phase equal to zero~$\xi_0=0$ at the particle location. The maximum Lorentz factor is then always given by $\gamma_{\rm max}=1+2\,a^2$ whatever the polarization of the wave. In order to check the integration of the particle equation of motion in a constant electromagnetic field, we compare the exact analytical solution with the constant field integrator. Several examples are shown without radiation reaction, a strength parameter up to $a=\numprint{e12}$ and circular or \LP modes such that $\alpha=\{0,0.5\}$.

\begin{figure}
	\centering
	\includegraphics[width=0.95\linewidth]{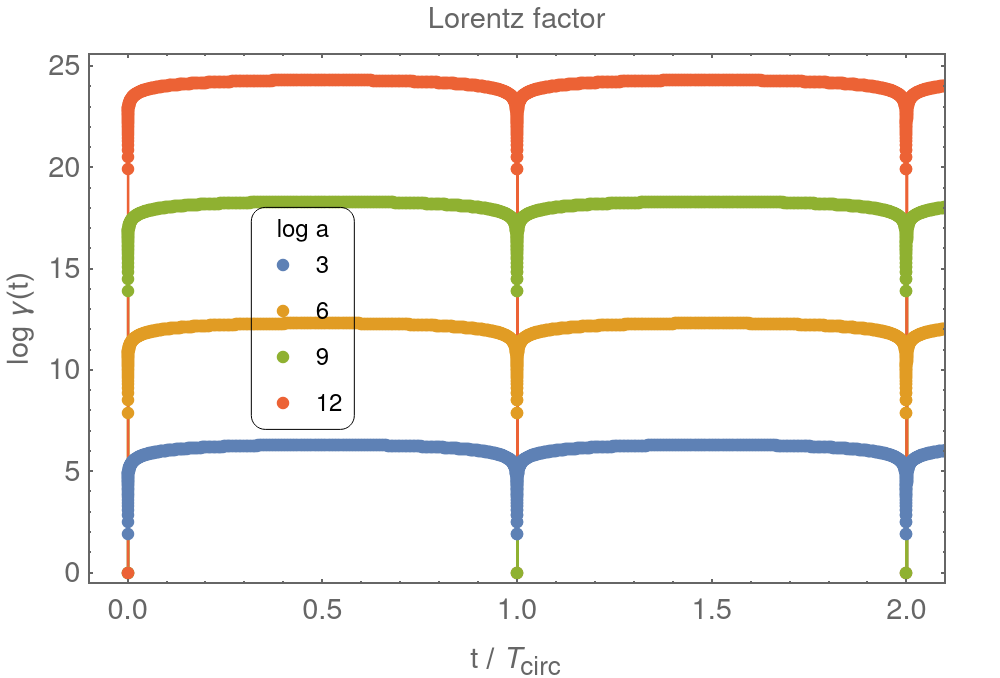}
	\caption{Evolution of the Lorentz factor of a particle initially at rest and for a circularly polarized wave with $\log a=\{3, 6, 9, 12\}$. Solid lines represent the exact analytical solutions and dotted points the constant field approximation.}
	\label{fig:lorentz_plane_e0.0_ax_b0_o1_p0_g0_mx}
\end{figure}

Fig.~\ref{fig:lorentz_plane_e0.0_ax_b0_o1_p0_g0_mx} shows the periodic variation of the Lorentz factor for a circularly polarized wave and a strength parameter $\log a=\{3, 6, 9, 12\}$. The time is normalised with respect to the period for a circularly polarised wave~$T_{\rm circ}$ given by Eq.~\eqref{eq:Periode_Circ} with $\beta=0$. The numerical solution marked as symbols perfectly overlaps with the analytical solution in solid line.

\begin{figure}
	\centering
	\includegraphics[width=0.95\linewidth]{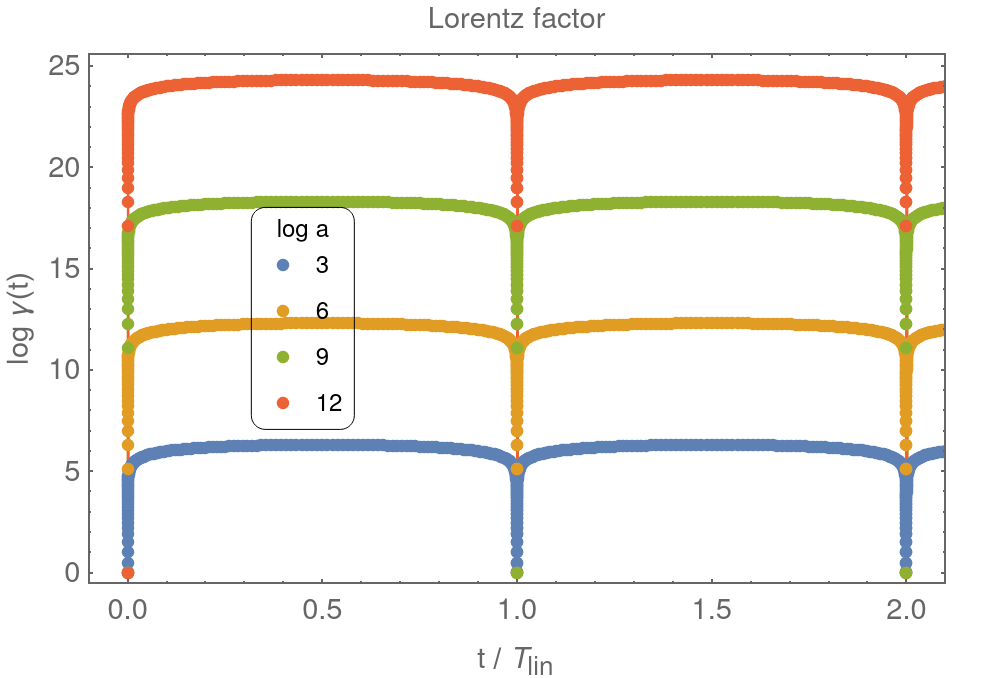}
	\caption{Same as Fig.~\ref{fig:lorentz_plane_e0.0_ax_b0_o1_p0_g0_mx} but for a linearly polarized wave.}
	\label{fig:lorentz_plane_e0.5_ax_b0_o1_p0_g0_mx}
\end{figure}

Fig.~\ref{fig:lorentz_plane_e0.5_ax_b0_o1_p0_g0_mx} shows the equivalent results for a linearly polarized wave and time normalisation according to~$T_{\rm lin}$ given by Eq.~\eqref{eq:Periode_Lin} with $\beta=0$. Here also, the match is perfect.

In this section, we saw that the particle gained energy from the wave but at the end of a cycle, i.e. after a phase variation~$\xi$ of $2\,\upi$, the particle returned to a state at rest, losing its kinetic energy due to the "braking" of the wave. The process is fully reversible in time for the Lorentz force. This is typical of a wave-particle interaction. We will see that when dissipation is added to the equation of motion, like for instance radiation reaction, the particle does not return to rest but keeps a minimal kinetic energy. The process is no longer fully time reversible.

\subsection{Particle starting at relativistic speed}

\begin{figure}
	\centering
	\includegraphics[width=0.95\linewidth]{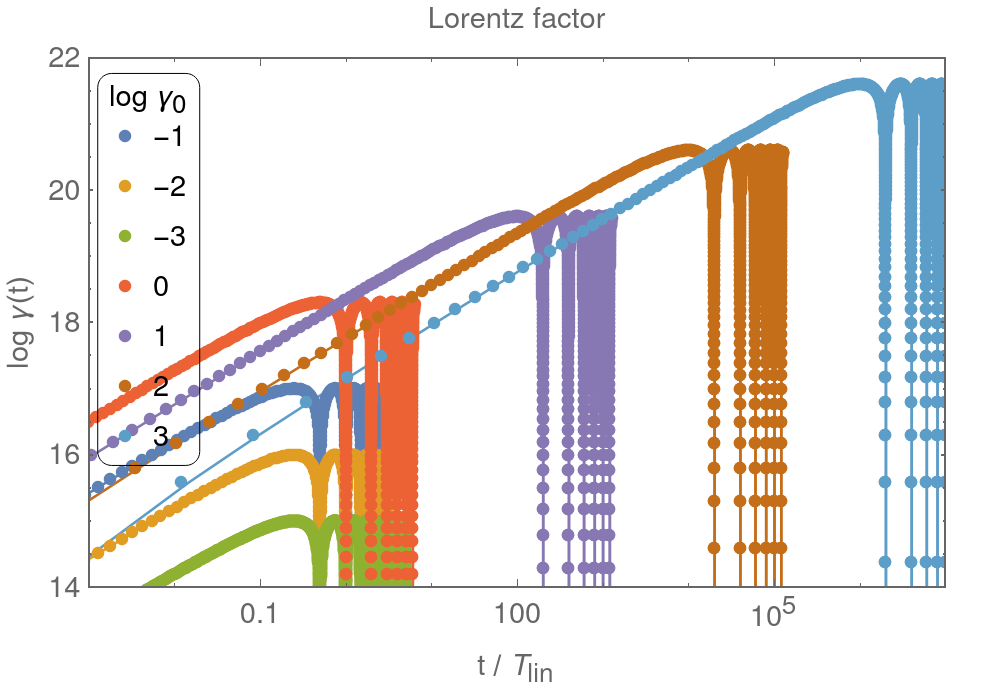}
	\caption{Evolution of the Lorentz factor of a particle injected with a relativistic speed and for a circularly polarized wave with $a=\numprint{e9}$. The legend shows $\log(\gamma_0)$ with the convention that a negative value means a velocity vector pointing in a direction opposite to the wave propagation.}
	\label{fig:lorentz_plane_e0.5_a9_b0_o1_p0_gx_mx}
\end{figure}

\begin{figure}
	\centering
	\includegraphics[width=0.95\linewidth]{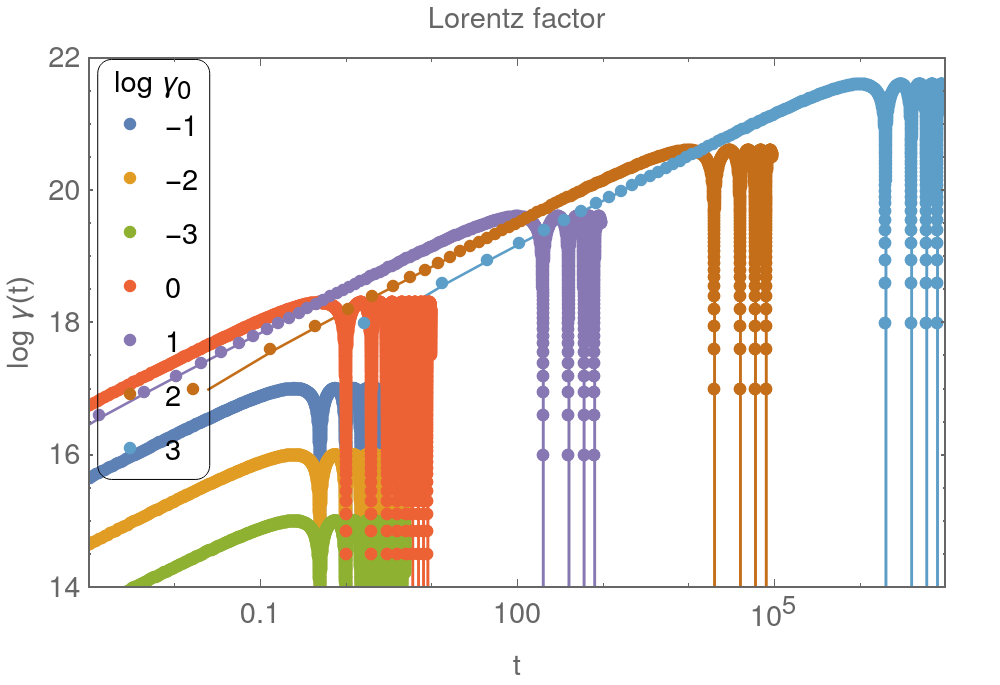}
	\caption{Same as Fig.~\ref{fig:lorentz_plane_e0.5_a9_b0_o1_p0_gx_mx} but for a linearly polarized wave.}
	\label{fig:lorentz_plane_e0.0_a9_b0_o1_p0_gx_mx}
\end{figure}

If the particle enters the wave with an initial relativistic velocity, the situation changes from the evolution found previously. A Lorentz boost in the rest frame of the particle does not affect the nature of the wave, it remains null-like but the wave frequency is Doppler shifted to a new frequency~$\omega'$ according to Eq.~\eqref{eq:Doppler_Frequence}. The periodicity in the  particle Lorentz factor also changes to Eq.~\eqref{eq:Periode_Circ} or to Eq.~\eqref{eq:Periode_Lin} depending on the wave polarization. Some examples for a circularly polarized wave are shown in Fig.~\ref{fig:lorentz_plane_e0.5_a9_b0_o1_p0_gx_mx} and for a linearly polarized wave in Fig.~\ref{fig:lorentz_plane_e0.0_a9_b0_o1_p0_gx_mx}. The initial Lorentz factor is $\gamma_0$ and shown in the legends as a logarithmic $\log\gamma_0$ with the convention that a negative value means a velocity vector pointing in a direction opposite to the wave propagation. The constant field approximation, in dotted points, agrees with the exact analytical solution, in solid lines.

\subsection{Initial phase of the wave}

\begin{figure}
	\centering
	\includegraphics[width=0.95\linewidth]{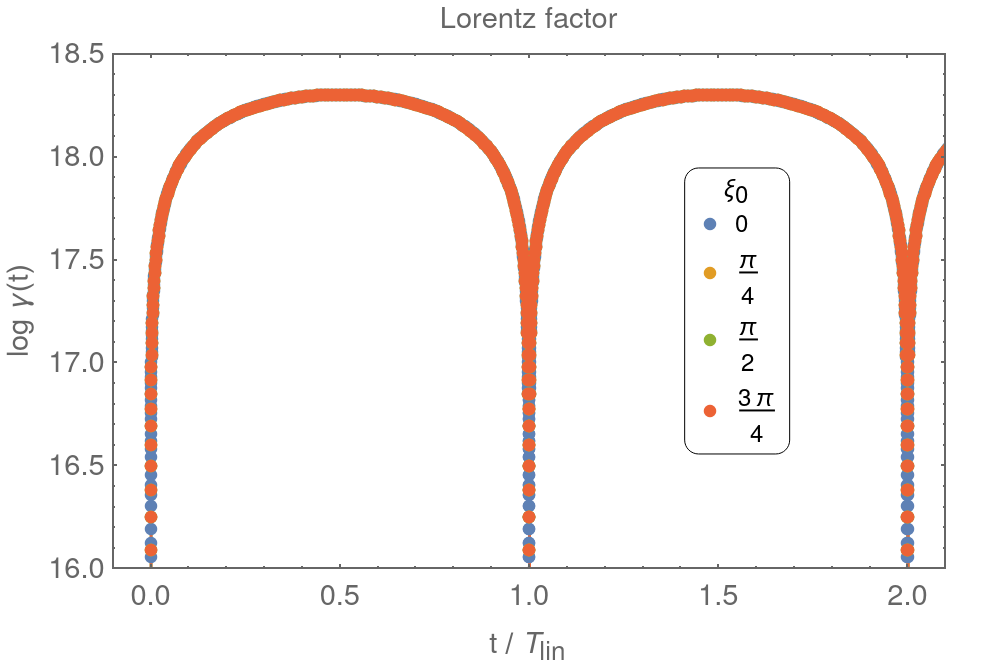}
	\caption{Evolution of the Lorentz factor of a particle injected at different initial phases for a circularly polarized wave with $a=\numprint{e9}$.}
	\label{fig:lorentz_plane_e0.0_a9_b0_o1_px_g0_mx}
\end{figure}

The phase~$\xi_0$ when the particle enters the wave also affects its subsequent trajectory. The impact of this initial phase is scrutinised by varying~$\xi_0$ in multiples of $\upi/4$ in the set $\xi_0 \in \{0, \upi/4, \upi/2, 3\,\upi/4\}$.
Some results are shown for a circularly polarized wave in Fig.~\ref{fig:lorentz_plane_e0.0_a9_b0_o1_px_g0_mx} by fixing the strength parameter to $a=\numprint{e9}$. As expected, for such waves, the trajectory is independent of the initial phase because only the $(\mathbf{E}, \mathbf{k})$ plane rotates without varying the strength of $\mathbf{E}$ or $\mathbf{B}$ with $\xi$. The maximum Lorentz factor is always $\gamma_{\rm max} = 1+2\,a^2 \approx \numprint{2e18}$.

\begin{figure}
	\centering
	\includegraphics[width=0.95\linewidth]{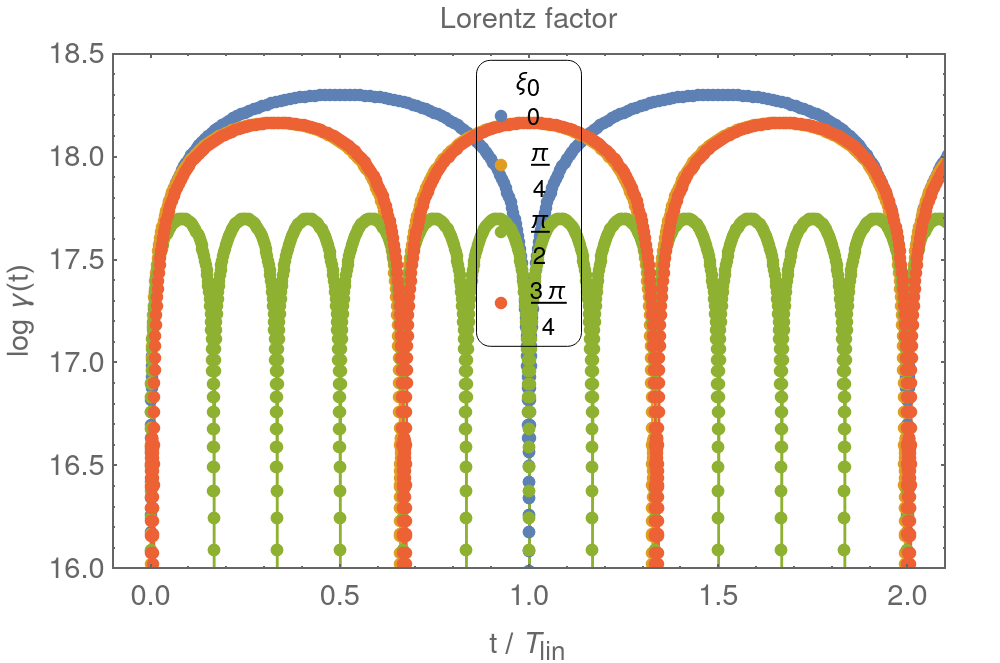}
	\caption{Same as Fig.~\ref{fig:lorentz_plane_e0.0_a9_b0_o1_px_g0_mx} but for a linearly polarized wave.}
	\label{fig:lorentz_plane_e0.5_a9_b0_o1_px_g0_mx}
\end{figure}

For linearly or elliptically polarized waves, the initial phase impacts the trajectory and the maximum Lorentz factor as shown for instance in Fig.~\ref{fig:lorentz_plane_e0.5_a9_b0_o1_px_g0_mx} for a linearly polarized wave with $a=\numprint{e9}$. Acceleration is most effective when injection happens at $\xi_0=0$ and a factor 4 less efficient if injection occurs at $\xi_0=\upi/2$.

This section demonstrated that the constant field approximation algorithm for null-like electromagnetic fields retrieves accurately the exact analytical motion of an ultra-relativistic particle in a elliptically polarized plane wave due to the Lorentz force, neglecting radiation reaction. Next we add the radiation feedback.

\section{Plane wave tests with radiation reaction}\label{sec:Tests2}

In conditions prevailing around rotating magnetized neutron stars, the electromagnetic field strength and the particle Lorentz factors are so large that radiation reaction efficiently slows down the particle by lowering its kinetic energy, converting it into radiation. In this section we redo the same analysis as in the previous section except that we add the radiation reaction without removing any term in the \LL approximation. 

The strength of radiation damping is controlled by the normalised parameter~$b$ defined in Eq.~\eqref{eq:Parametres}. Typical values for neutron stars are
\begin{equation}\label{eq:damping_b}
 b = \numprint{4e-23} \, \left( \frac{P}{1~\SIunits{\second}}\right)^{-1}
\end{equation}
where $P=2\,\upi/\Omega$ is the pulsar period in second.
It is strongest for millisecond pulsars reaching values of $b \approx \numprint{2e-20}$ for a 2~ms pulsar. 
For the simulations presented below, we use $\log b=-20$. 
The perturbation in the Lorentz force also includes terms involving $\gamma$ and $a$ as explained in the paragraph after Eq.~\eqref{eq:LLnormee}.

\subsection{Particle starting at rest}

\begin{figure}
	\centering
	\includegraphics[width=0.95\linewidth]{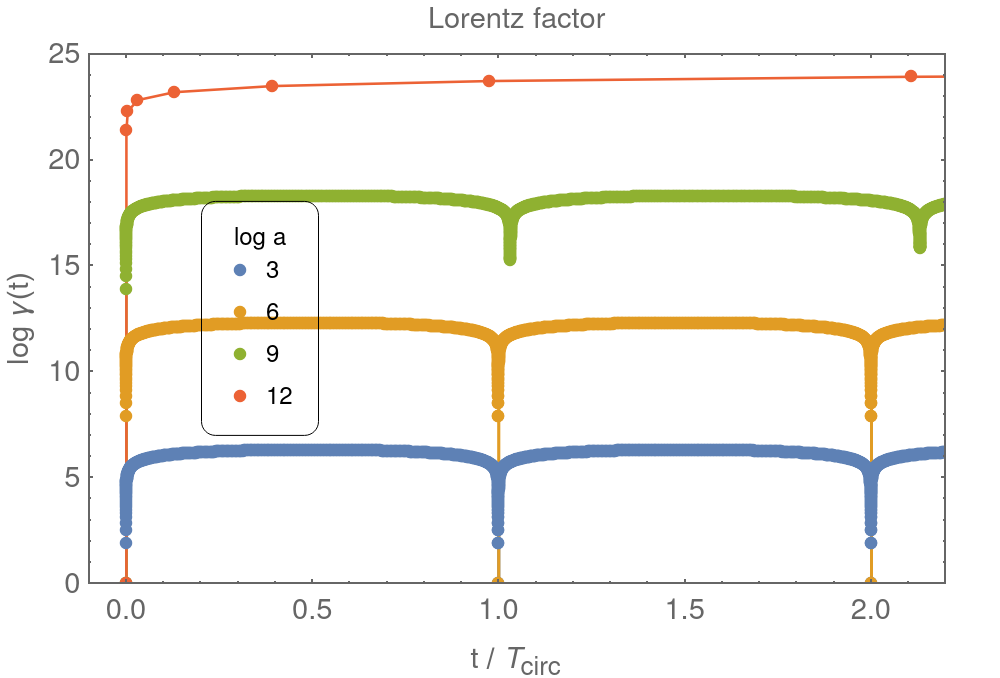}
	\caption{Evolution of the Lorentz factor of a particle initially at rest with radiation reaction set to $\log b=-20$ and for a circularly polarized wave with $\log a=\{3, 6, 9, 12\}$.}
	\label{fig:lorentz_plane_e0.0_ax_bx_o1_p0_g0_mx}
\end{figure}

\begin{figure}
	\centering
	\includegraphics[width=0.95\linewidth]{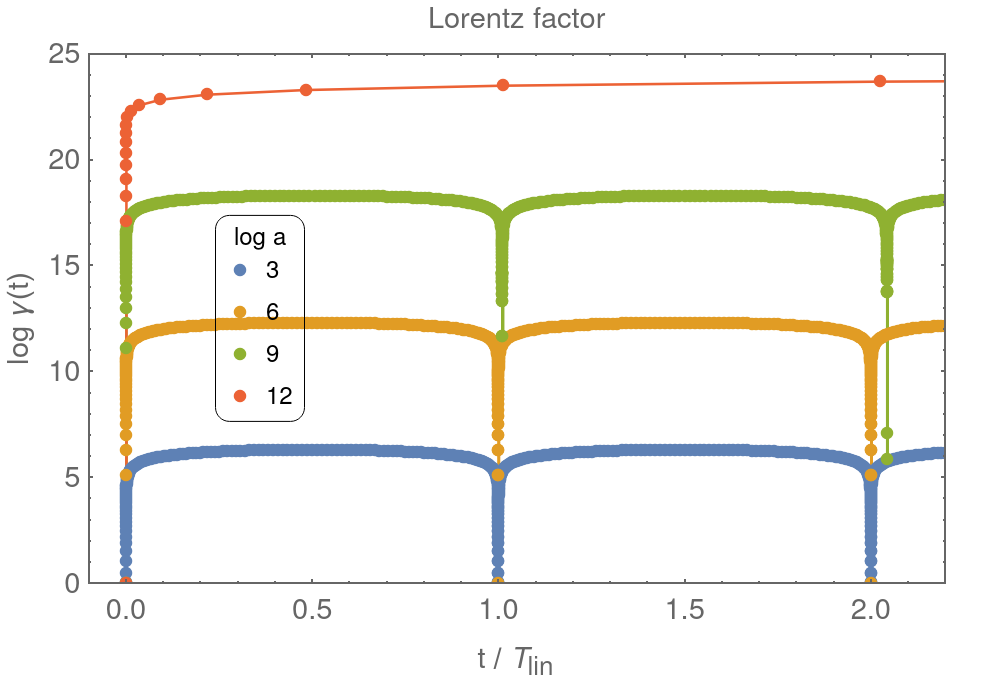}
	\caption{Same as Fig.~\ref{fig:lorentz_plane_e0.0_ax_bx_o1_p0_g0_mx} but for a linearly polarized wave.}
	\label{fig:lorentz_plane_e0.5_ax_bx_o1_p0_g0_mx}
\end{figure}

When the particle starts at rest, the radiation reaction vanishes. Whatever the strength and damping parameters $a$ and $b$, the particle evolves initially only according to the Lorentz force. The trajectories are therefore identical to the previous cases without radiation reaction. Only when the Lorentz factor reaches high enough values for the perturbation to become to the same order of magnitude as the Lorentz part will the particle deviate from its dissipationless motion. This is seen in Fig.~\ref{fig:lorentz_plane_e0.0_ax_bx_o1_p0_g0_mx} showing the particle Lorentz factor evolving in a circularly polarized wave for $\log a=\{3, 6, 9, 12\}$. By inspection of Fig.~\ref{fig:lorentz_plane_e0.5_ax_bx_o1_p0_g0_mx}, we deduce that the behaviour in a linearly polarized wave is very similar, only the largest strength parameters leading to the largest Lorentz factors will perturb the Lorentz force. Indeed, only the case $a=\numprint{e12}$ leads to the radiation dominated motion in the regime $a^2\,b = \numprint{e4} \gg 1$. All other cases a well approximated by the Lorentz force motion, except for $a=\numprint{e9}$ where we observe a slight increase in the periodic variation in $\gamma$ with time, see the plots in green point in Fig.~\ref{fig:lorentz_plane_e0.0_ax_bx_o1_p0_g0_mx} and Fig.~\ref{fig:lorentz_plane_e0.5_ax_bx_o1_p0_g0_mx}.

\begin{figure}
	\centering
	\includegraphics[width=0.95\linewidth]{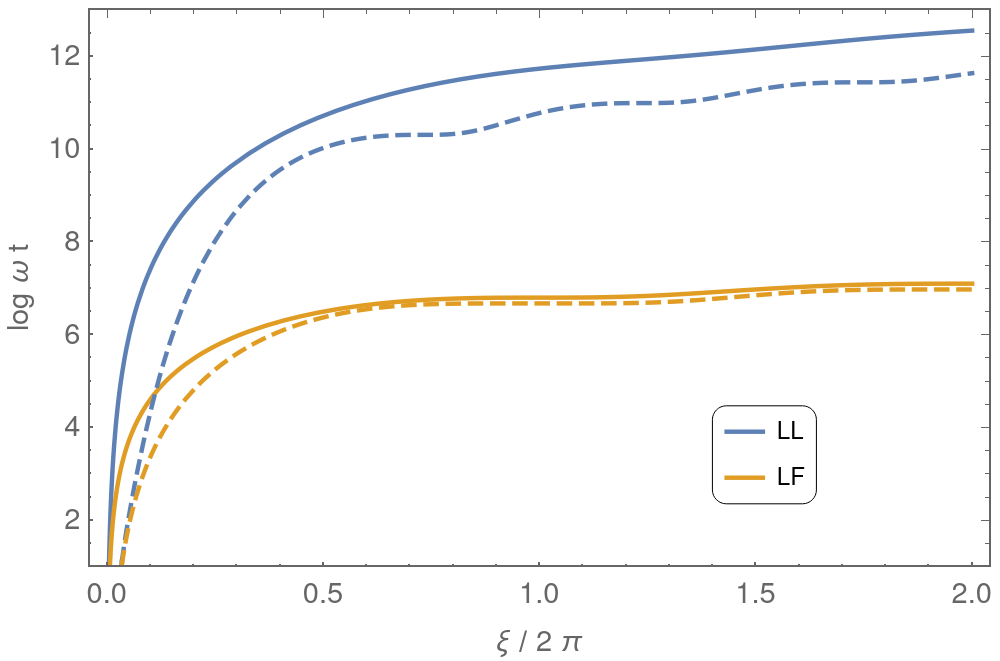}
	\caption{Phase evolution of the inertial frame clock normalized time $\omega\,t$ of a particle initially at rest with and without radiation reaction (resp. LL and LF) for a circularly (solid line) and linearly (dashed line) polarized wave with $\log a=3$ and $\log b=-4$.}
	\label{fig:lorentz_plane_RR_temps}
\end{figure}

\begin{figure}
	\centering
	\includegraphics[width=0.95\linewidth]{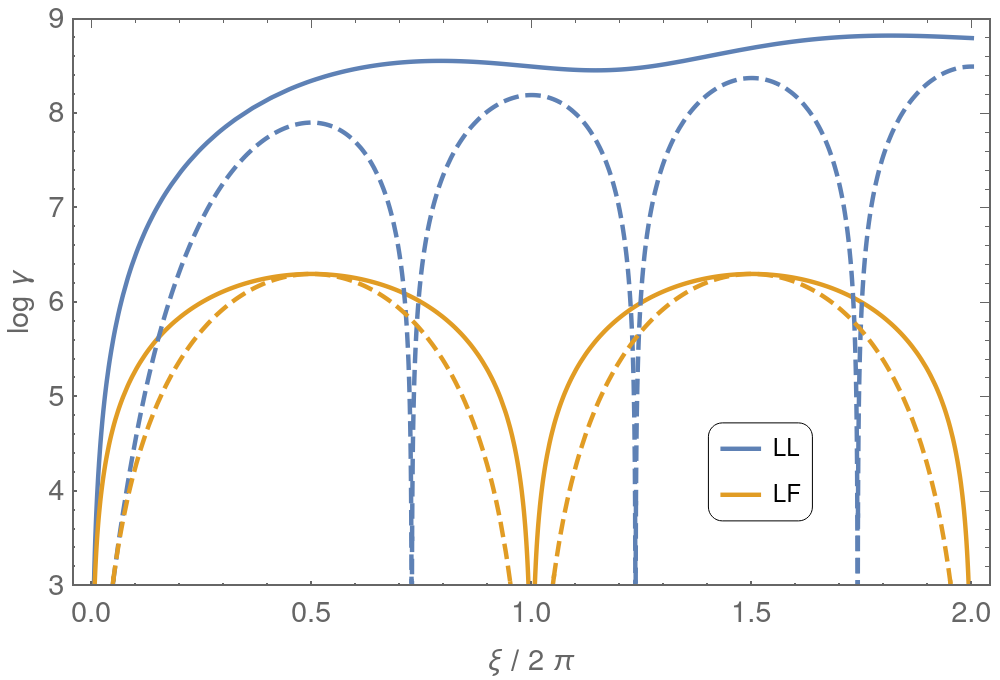}
	\caption{Same as Fig~\ref{fig:lorentz_plane_RR_temps} but for the phase evolution of the Lorentz factor.}
	\label{fig:lorentz_plane_RR}
\end{figure}

\begin{figure}
	\centering
	\includegraphics[width=0.95\linewidth]{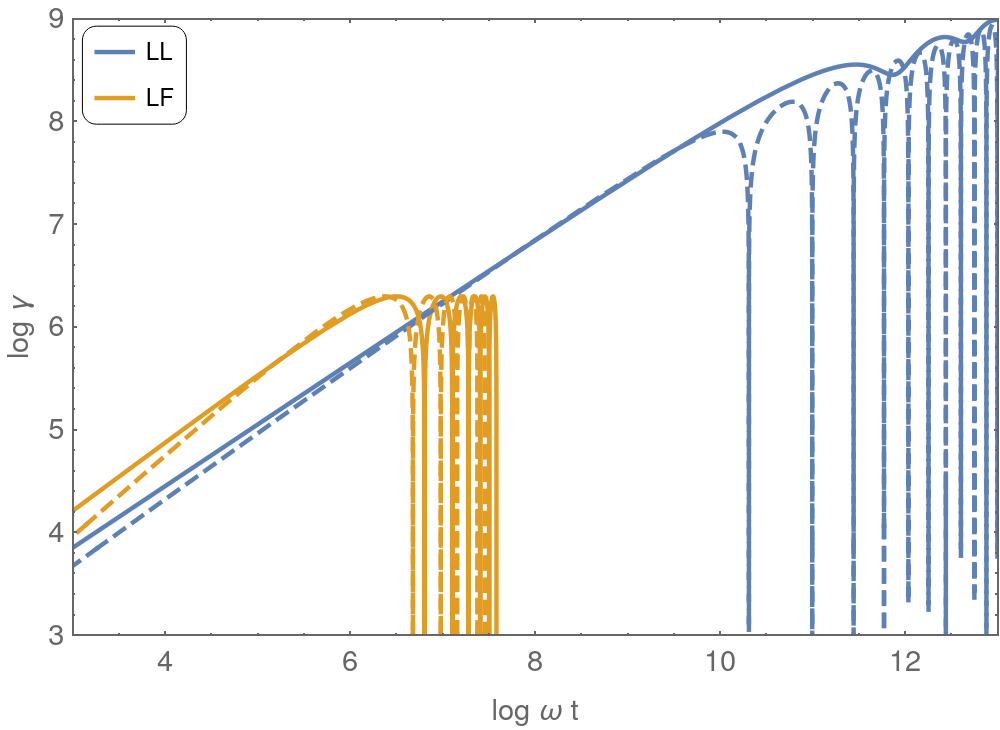}
	\caption{Evolution of the Lorentz factor with distance according to Fig.~\ref{fig:lorentz_plane_RR_temps} and Fig.~\ref{fig:lorentz_plane_RR}.}
	\label{fig:lorentz_plane_RR_r_gamma}
\end{figure}

Radiation reaction drastically inflates the typical time scale of Lorentz factor variation as can be checked in Fig.~\ref{fig:lorentz_plane_RR_temps} showing an increase by 10 orders of magnitude in the case of $a=\numprint{e12}$ for circularly as well as for linearly polarized waves, respectively in solid lines and dashed lines with and without radiation reaction (resp. LL in blue and LF in orange).
The maximum Lorentz factor also increases significantly when radiation reaction is included, see Fig.~\ref{fig:lorentz_plane_RR}. In the aforementioned case, there is an increase by 4 to 5 orders of magnitude.
Finally, Fig.~\ref{fig:lorentz_plane_RR_r_gamma} summarizes the spatial evolution of this Lorentz factor, demonstrating the stretching effect of radiation reaction. The achievable energy is much higher but it requires more time or space to attain its asymptotic value.


We conclude that when the radiation reaction is taken into account, i.e., in the presence of deceleration induced by radiative friction, the particle energy becomes greater than without taking it into account. This statement appears counter-intuitive but it is not related to the well-known runaway solutions of the LAD equation because these motions do not show any exponential grow of the Lorentz factor as would be the case for a runaway solution. Indeed, the Landau \& Lifshits prescription is free of these parasitic solutions because it is a second order in time equation of motion. Therefore the non-physical self-accelerating solutions are absent in the \LL equation. The reason leading to more efficient acceleration in case of radiation reaction is related to the precise time evolution of the particle in the electromagnetic wave.

\cite{gunn_motion_1971} showed indeed that, for any initial conditions, a particle evolving in a plane electromagnetic wave with radiation reaction (in the \LL prescription) slowly increases its energy with time, a kind of "radiative pumping" as they said. The radiation reaction can be interpreted as a friction causing a delay in the particle response to the field, inducing a lag between its velocity and the accelerating electric field, causing the slowly in time increase in kinetic energy, typically $\gamma \propto t^{1/3}$ as they showed. The fact that radiation reaction can decrease or increase the Lorentz factor in plane waves has also been noticed by \cite{heintzmann_acceleration_1972}. Such pumping is not effective in spherical waves because the kinetic energy increase occurs mainly during the phase locked motion at the beginning of the acceleration process and requires many cycles with constant strength parameter.

\subsection{Particle starting at relativistic speed}

\begin{figure}
	\centering
	\includegraphics[width=0.95\linewidth]{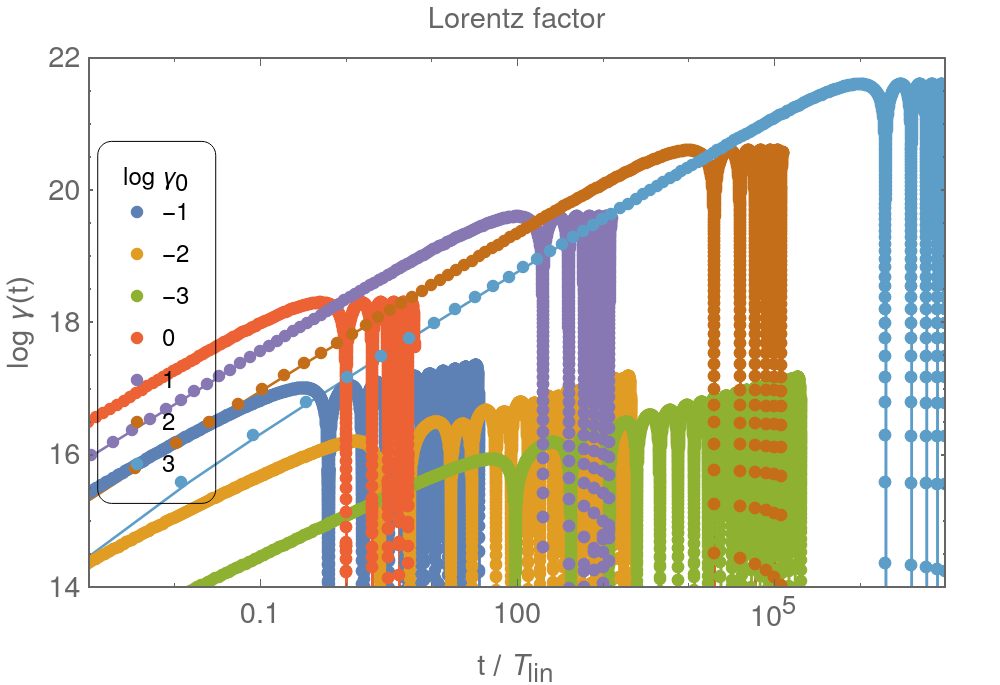}
	\caption{Evolution of the Lorentz factor of a particle injected with a relativistic speed and for a circularly polarized wave with $a=\numprint{e9}$.}
	\label{fig:lorentz_plane_e0.5_a9_bx_o1_p0_gx_mx}
\end{figure}

\begin{figure}
	\centering
	\includegraphics[width=0.95\linewidth]{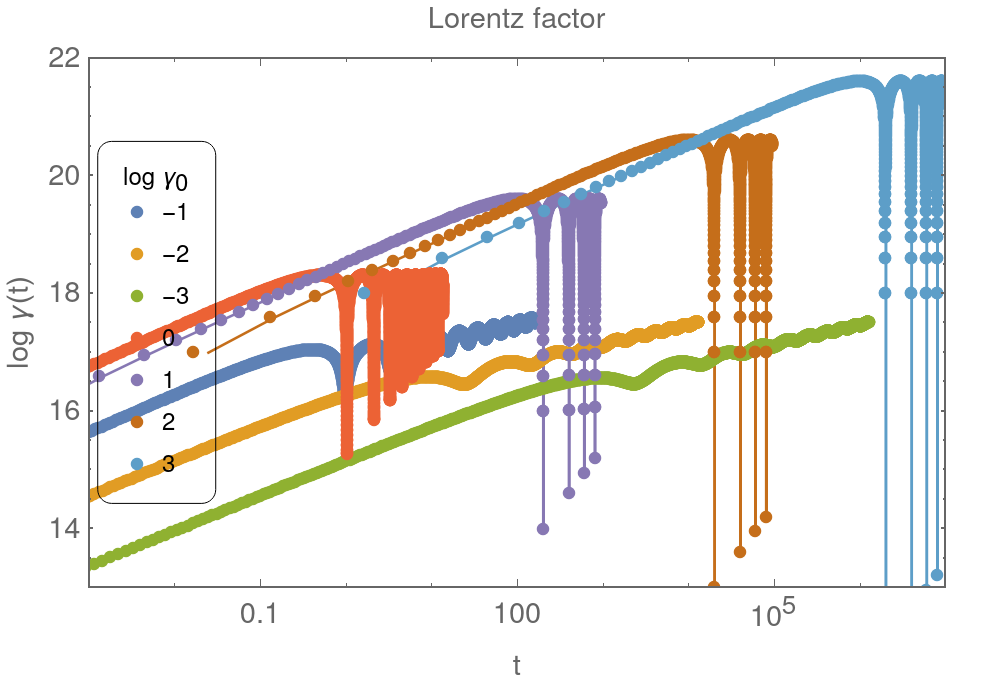}
	\caption{Same as Fig.~\ref{fig:lorentz_plane_e0.5_a9_bx_o1_p0_gx_mx} but for a linearly polarized wave.}
	\label{fig:lorentz_plane_e0.0_a9_bx_o1_p0_gx_mx}
\end{figure}

When the particle starts at a relativistic speed, for the same simulation runs as in the previous section, the maximum Lorentz factor reached by the particle is not sufficient to significantly perturb the Lorentz force if the particle catches up the wave. We therefore do not observe any difference between radiation reaction and solely Lorentz force evolution when inspecting Fig.~\ref{fig:lorentz_plane_e0.5_a9_bx_o1_p0_gx_mx} for a circularly polarized wave or Fig.~\ref{fig:lorentz_plane_e0.0_a9_bx_o1_p0_gx_mx} for a linearly polarized wave, in the cases marked with a positive $\log \gamma_0$, meaning particles moving in the same direction as the wave. This is due to the fact that the effective damping parameter $b' = \omega' \, \tau_0$ as measured in the particle rest frame decreases by several orders of magnitude due to Doppler shifting of the wave frequency $\omega = \mathcal{D} \, \omega' \gg \omega'$. To the contrary, for a head on collision between the particle and the wave, the apparent wave frequency is blue shifted due to the Doppler effect, and the effective damping parameter $b'$ increases by several orders of magnitude. Radiation reaction becomes significant and the particle trajectory is affected by the perturbing force. This is clearly seen for negative $\log\gamma_0$ (meaning particle moving in opposite direction to the wave propagation i.e. a head-on collision) in Fig.~\ref{fig:lorentz_plane_e0.5_a9_bx_o1_p0_gx_mx} and Fig.~\ref{fig:lorentz_plane_e0.0_a9_bx_o1_p0_gx_mx} where the Lorentz factor slowly drifts to larger and larger values.

\subsection{Initial phase of the wave}

\begin{figure}
	\centering
	\includegraphics[width=0.95\linewidth]{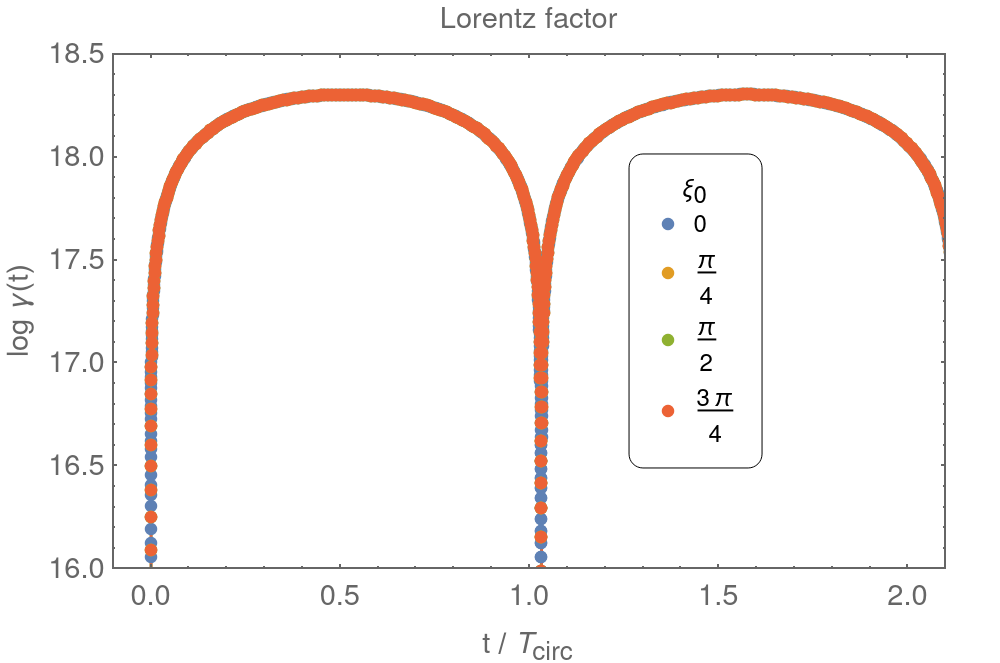}
	\caption{Evolution of the Lorentz factor of a particle injected at different initial phases for a circularly polarized wave with $a=\numprint{e9}$.}
	\label{fig:lorentz_plane_e0.0_a9_bx_o1_px_g0_mx}
\end{figure}

\begin{figure}
	\centering
	\includegraphics[width=0.95\linewidth]{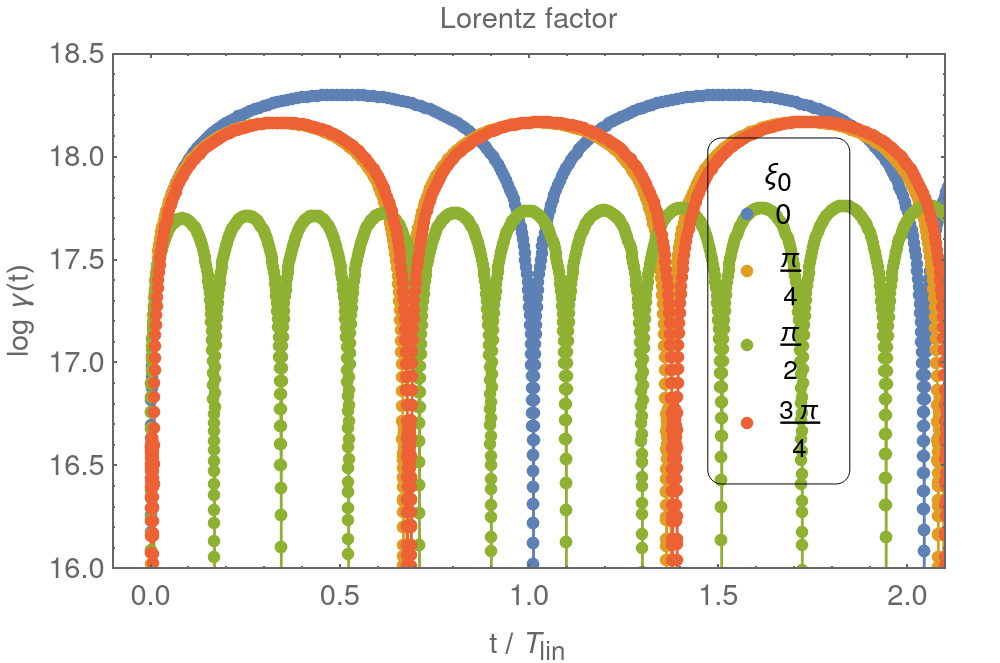}
	\caption{Same as Fig.~\ref{fig:lorentz_plane_e0.0_a9_bx_o1_px_g0_mx} but for a linearly polarized wave.}
	\label{fig:lorentz_plane_e0.5_a9_bx_o1_px_g0_mx}
\end{figure}

We already saw that a circularly polarized wave is insensitive to the initial phase due to its symmetry of rotation about an axis parallel to the wave vector $\mathbf{k}$. This holds true for radiation reaction as demonstrated in Fig.~\ref{fig:lorentz_plane_e0.0_a9_bx_o1_px_g0_mx} and as expected due to this symmetry property.
However, as for the Lorentz force, the \LL equation remains also sensitive to the initial phase for the linearly polarized wave as seen in Fig.~\ref{fig:lorentz_plane_e0.5_a9_bx_o1_px_g0_mx}.

We recall that in all the above exposed examples, the exact analytical solutions for the 4-velocity are known and served as a check for our algorithm. We found that exact analytical solutions for the constant field approximation gives sensibly the same results. Therefore, it demonstrates that the constant field approximation can serve as a building block for very general null-like electromagnetic fields.

After these extensive tests of our numerical algorithm for null-like electromagnetic fields, we apply it in the context of neutron star vacuum magnetospheres outside the light-cylinder where spherical waves are launched, described locally as plane waves to a good approximation.

\section{Spherical wave results}\label{sec:Results}

Since the work of \cite{deutsch_electromagnetic_1955}, we know that a rotating magnetic dipole launches a large amplitude low frequency electromagnetic wave at large distances $r\gg\rlight$, that is approximated by a spherical wave of definite polarization depending on the colatitude~$\theta$ and decreasing with distance like $\rlight/r$. Indeed, along, the rotation axis, the wave is a circularly polarized wave whereas along the equator, it is completely linearly polarized. In between these two limits, the wave shows any degree of elliptical polarization, left handed or right handed.

In this last section, we apply our algorithm to a real astrophysical context of particle acceleration and radiation in the wave zone outside the light-cylinder of a neutron star. As a typical value of the magnetic field strength at this light-cylinder, we choose $B_{\rm L}\approx \numprint{1}~\SIunits{\tesla}$. The light-cylinder is also the place where the quasi-static regime transit to the wave zone. By default, we assume that particles enter the wave at a radius~$r_0$ equal to the light-cylinder if not otherwise specified, $r_0=\rlight$. Moreover, we employ the spherical wave approximation meaning a decrease in the field amplitude like $(E,B) \propto \rlight/r$ and where the wave field components $\mathbf{E}$, $\mathbf{B}$ and propagation direction $\mathbf{n}$ are mutually orthogonal. Any kind of polarization can be considered, linear, left/right circular and elliptical polarization.

\subsection{Particle starting at rest}

Let us assume that particles are injected at rest at a distance $r_0=\rlight$ from the neutron star centre and an arbitrary colatitude~$\theta$ with respect to the rotation axis. Because the wave amplitude decrease with distance, particles do not reach the maximum energy $\gamma_{\rm max}$ of a plane wave. The actual maximum energy is much less and does not scale as $(1+2\,a^2)$ any more as we will proof.

\begin{figure}
	\centering
	\includegraphics[width=0.95\linewidth]{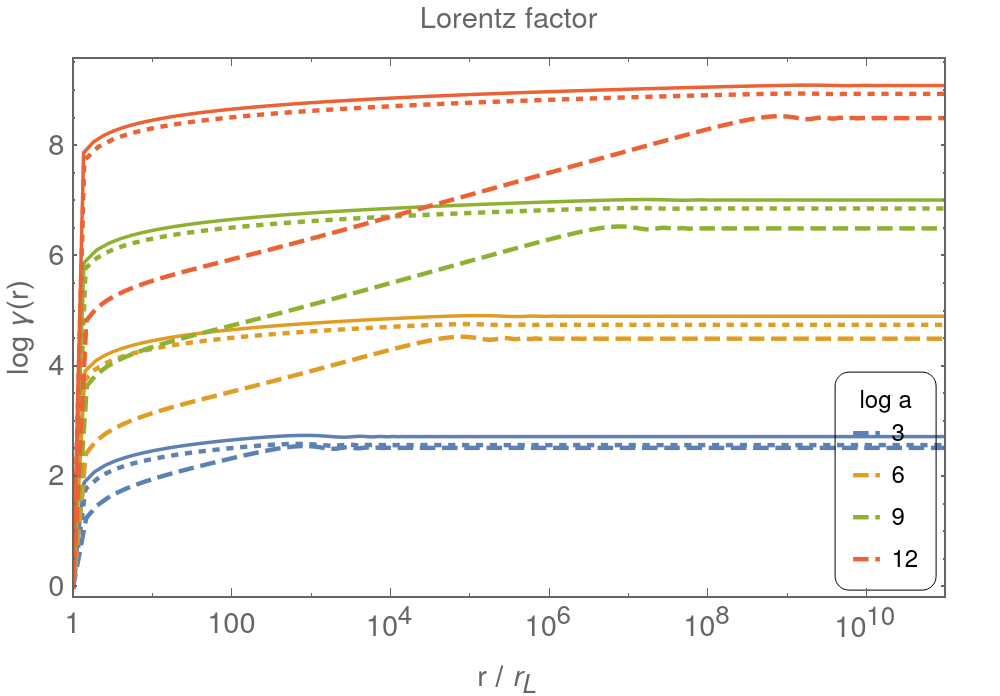}
	\caption{Evolution of the Lorentz factor for a circularly polarized wave ($\alpha=0$) in solid line, an elliptically  polarized wave ($\alpha=0.2$) in dotted line and linearly polarized wave ($\alpha=0.5$) in dashed line. The strength parameter is $\log a=\{3,6,9,12\}$.}
	\label{fig:lorentz_plane_ex_ax_b0_o2_p0_g0_m1}
\end{figure}

\begin{figure}
	\centering
	\includegraphics[width=0.95\linewidth]{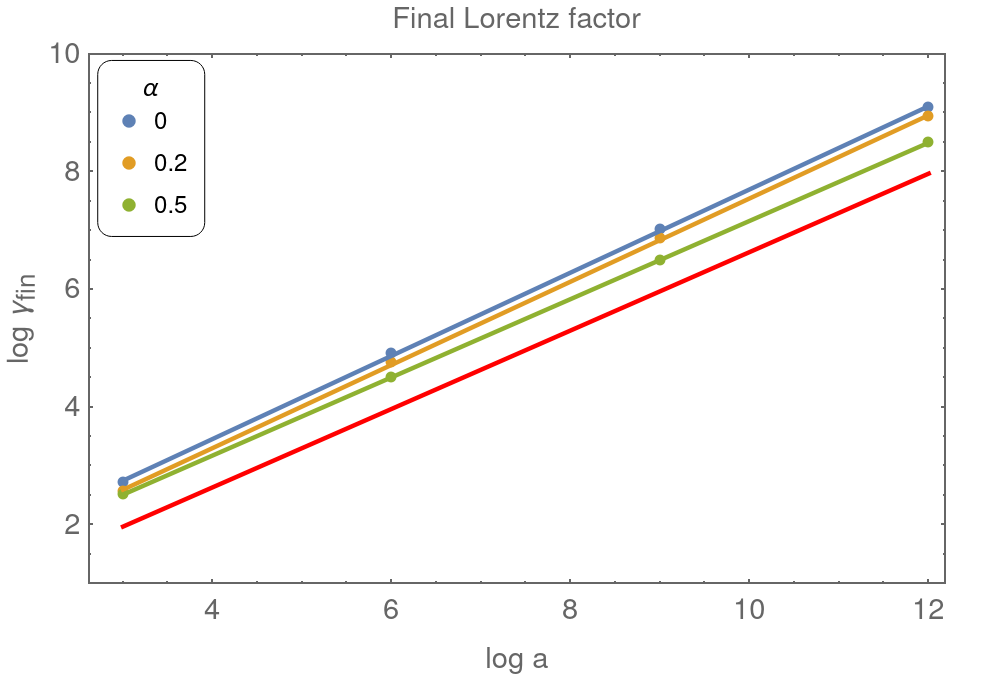}
	\caption{Final Lorentz factor for a circularly ($\alpha=0$), elliptically ($\alpha=0.2$) and linearly ($\alpha=0.5$) polarized wave. The law in Eq.~\eqref{eq:gamma_fin} is shown in red solid line.}
\label{fig:lorentz_final_ex_ax_b0_o2_p0_g0_m1}
\end{figure}

Indeed, Fig.~\ref{fig:lorentz_plane_ex_ax_b0_o2_p0_g0_m1} shows the acceleration efficiency for circularly, elliptically and linearly polarized waves respectively in solid, dotted and dashed line, with $\alpha =\{0,0.2,0.5\}$ without radiation reaction and strength parameter $\log a=\{3,6,9,12\}$. The maximum Lorentz factor found from these runs scales roughly as $a^{0.7}$ in all cases, the weakest values being obtained for a linear polarization, Fig.~\ref{fig:lorentz_final_ex_ax_b0_o2_p0_g0_m1}. Following the arguments exposed by \cite{michel_electrodynamics_1999}, the particle reaches its maximum energy after travelling a distance $r_{\rm c} \approx \upi\,a_{\rm c}^2$ where $a_{\rm c}$ is the strength parameter at the distance~$r_{\rm c}$. But at these distances, the strength parameter has decreased to a value $a_{\rm c} = a \, \rlight/r_{\rm c}$. Solving for the distance, we get $r_{\rm c} \approx \upi^{1/3} \, a^{2/3} \, \rlight$. A good guess of this final Lorentz factor is given by
\begin{equation}\label{eq:gamma_fin}
\gamma_{\rm fin} \approx 2 \, a_{\rm c}^2 \approx 2 \, (a / \upi)^{2/3}
\end{equation}
which is in agreement with the fitted exponent of $0.7\approx 2/3$. The law \eqref{eq:gamma_fin} is shown in red solid line in Fig.~\ref{fig:lorentz_final_ex_ax_b0_o2_p0_g0_m1}. The distance required to attain this asymptotic value is however much larger for the \LP ($\alpha=0.5$) compared to the circular or elliptic polarization with $\alpha=\{0,0.2\}$. Therefore, around a rotating magnet, acceleration is most effective along the rotation axis and weakest around the rotational equator.

\begin{figure}
	\centering
	\includegraphics[width=0.95\linewidth]{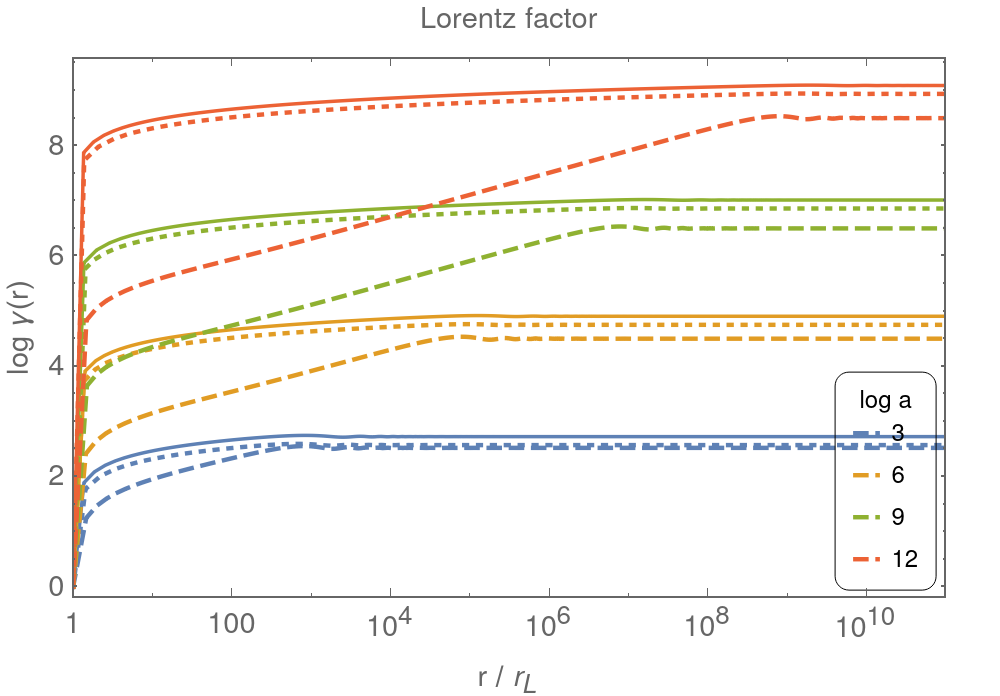}
	\caption{Same as Fig.~\ref{fig:lorentz_plane_ex_ax_b0_o2_p0_g0_m1} but with radiation reaction fixed to $\log b=-20$.}
	\label{fig:lorentz_plane_ex_ax_bx_o2_p0_g0_mx}
\end{figure}

Fig.~\ref{fig:lorentz_plane_ex_ax_bx_o2_p0_g0_mx} shows the same results as in Fig.~\ref{fig:lorentz_plane_ex_ax_b0_o2_p0_g0_m1} but with radiation reaction fixed to $\log b=-20$. Because the perturbation scales as $a^2\,b$ and $a$ decreases with distance, radiation feedback does not produce any significant perturbation to the particle motion, except in the efficient acceleration zone around the light-cylinder for the case $a=\numprint{e12}$. We conclude that radiation reaction does not impact the particle motion in the wave zone of a pulsar in this vacuum case.

\begin{figure}
	\centering
	\includegraphics[width=0.95\linewidth]{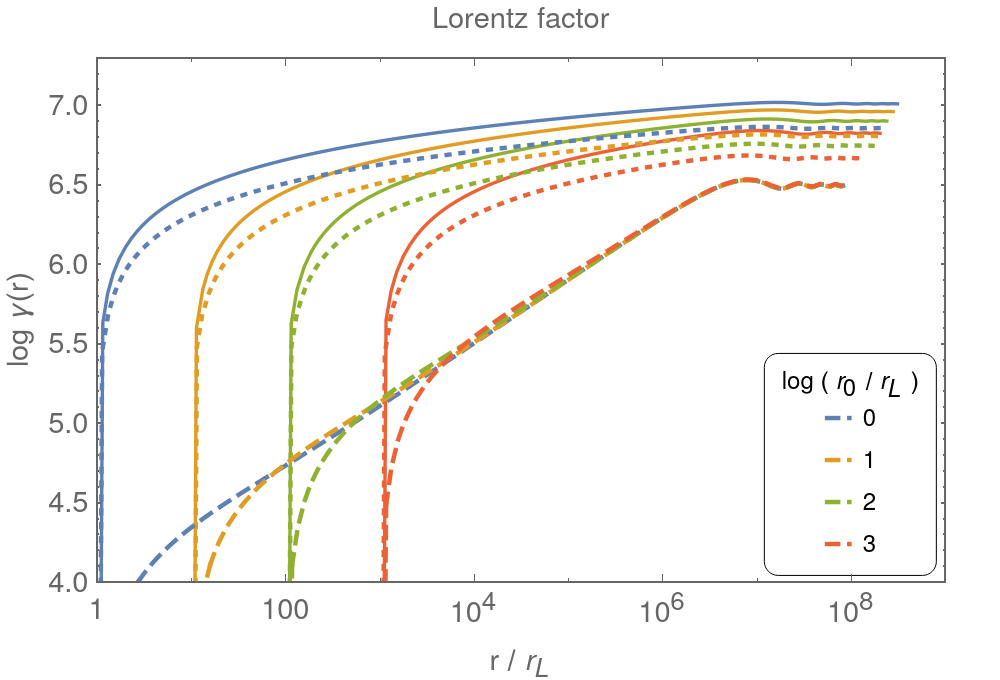}
	\caption{Evolution of the Lorentz factor with distance for a circularly ($\alpha=0$), elliptically ($\alpha=0.2$) and linearly ($\alpha=0.5$) polarized wave respectively in solid, dotted and dashed line for different initial positions $r_0 / \rlight = \{1,10,100,1000\}$ and with $a=10^9$.}
	\label{fig:lorentz_plane_ex_a9_b0_o2_p0_g0_m1}
\end{figure}

\begin{figure}
	\centering
	\includegraphics[width=0.95\linewidth]{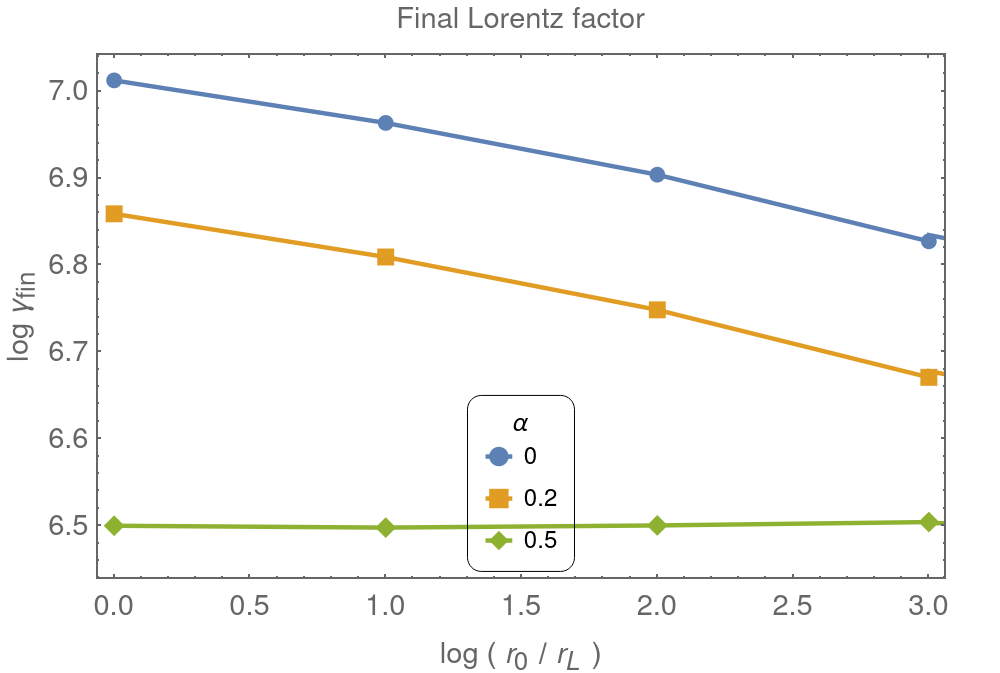}
	\caption{Maximum Lorentz factor, extracted from Fig.~\ref{fig:lorentz_plane_ex_a9_b0_o2_p0_g0_m1}, for a circularly ($\alpha=0$), elliptically ($\alpha=0.2$) and linearly ($\alpha=0.5$) polarized wave respectively in blue, orange and green line for different initial positions $\log(r_0 / \rlight) = \{0,1,2,3\}$ and $a=10^9$.}
	\label{fig:lorentz_final_rx_ex_a9_b0_o2_p0_g0_m1}
\end{figure}

The final Lorentz factor is also relatively insensitive to the initial position~$r_0$ of the particle at rest. Fig.~\ref{fig:lorentz_plane_ex_a9_b0_o2_p0_g0_m1} indeed shows the Lorentz factor dependence on distance for particles evolving in a spherical wave with $a=\numprint{e9}$, for several polarization states and several initial positions $\log (r_0 / \rlight) = \{0,1,2,3\}$.
For \LP as long as particles are injected at radii shorter than the distance where the asymptotic energy is attained, in the example about $r_{\rm c}/\rlight \approx \numprint{e7}$, the maximum Lorentz factor is noticeably the same. For \CP and elliptic polarization, this maximum $\gamma$ slightly decreases with $r_0$, not even by a factor two for a distance increase of three orders of magnitude, see Fig.~\ref{fig:lorentz_final_rx_ex_a9_b0_o2_p0_g0_m1}. The energy acquired by a particle therefore only depends on the wave characteristic, that is its polarization~$\alpha$ and strength parameter~$a$ for injections at distances $r<r_{\rm c}$. At distance $r>r_{\rm c}$, particle acceleration efficiency sharply decreases.

\subsection{Particle starting at relativistic speed}

Around a neutron star, particles entering the waves are injected already at high Lorentz factors from the magnetosphere, within the light-cylinder. We do not expect them to be picked up at rest by the wave, see the discussion below in Sec.~\ref{sec:Discussion}. Therefore we imposed initial conditions where particles catch up the wave at relativistic speed. We already saw that particles reach Lorentz factors well above $\gamma_{\rm max}$ imposed by the strength parameter~$a$ if the particle catches up the wave without radiation reaction. This scaling with $a^2$ is typical of a coherent wave/particle interaction in the phase-locking stage.

\begin{figure}
	\centering
	\includegraphics[width=0.95\linewidth]{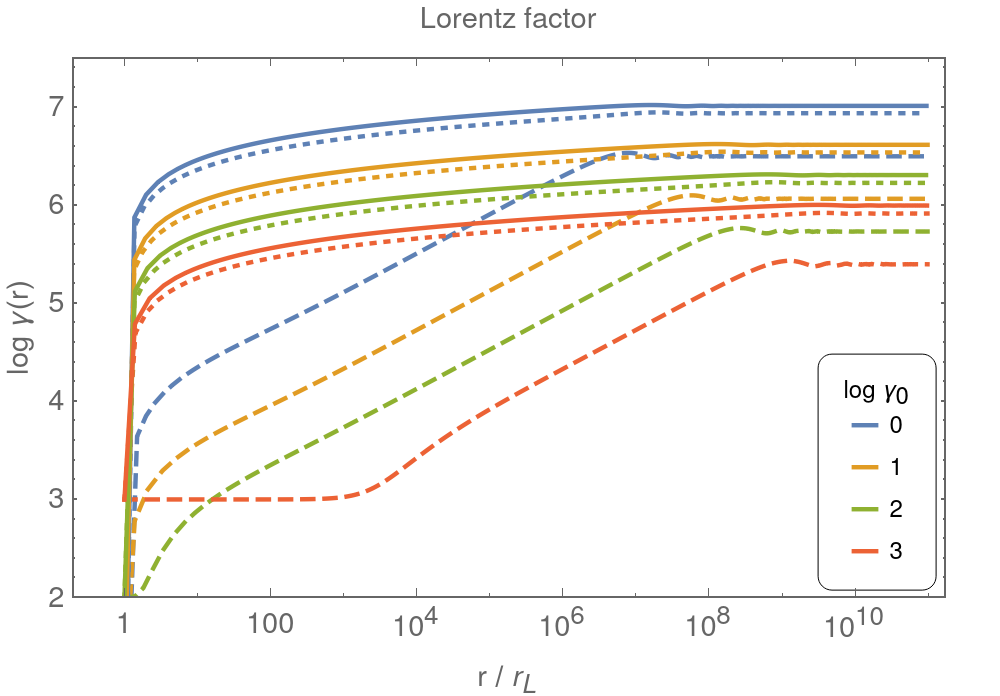}
	\caption{Evolution of the Lorentz factor for a linearly polarized wave ($\alpha=0.5$) with $a=\numprint{e9}$, initial Lorentz factor $\log \gamma_0=\{0,1,2,3\}$ and varying initial phases, $\xi_0=0$ in dashed lines, $\xi_0=\upi/4$ in dotted lines and $\xi_0=\upi/2$ in solid lines.}
	\label{fig:lorentz_plane_e0.5_a9_b0_o2_p0_gx_mx}
\end{figure}

\begin{figure}
	\centering
	\includegraphics[width=0.95\linewidth]{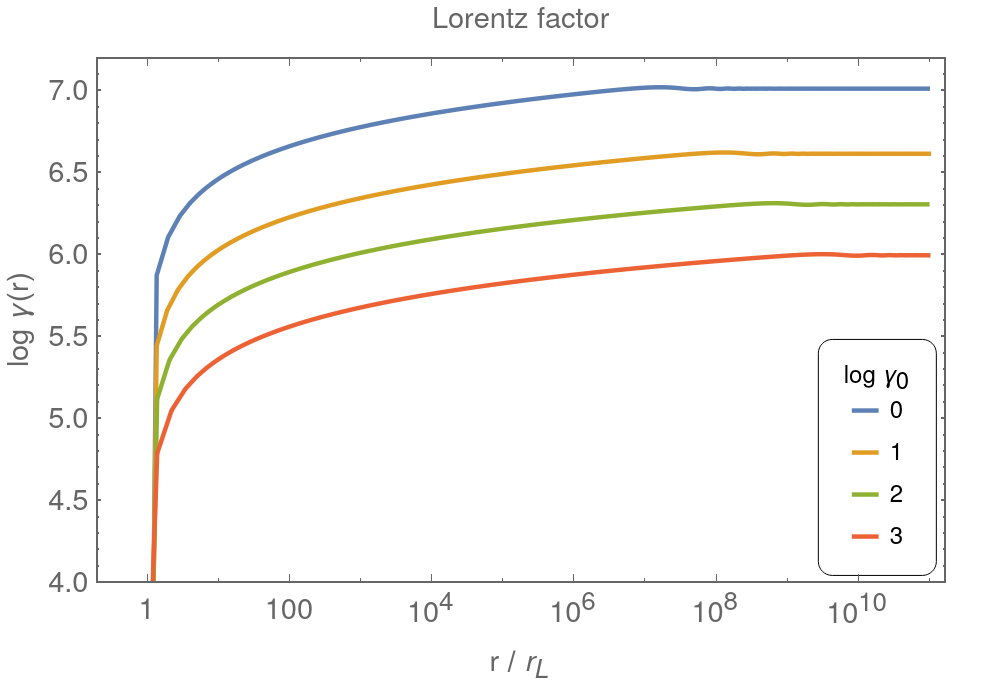}
	\caption{Evolution of the Lorentz factor for a circularly polarized wave with $a=\numprint{e9}$, initial Lorentz factor $\log \gamma_0=\{0,1,2,3\}$. The curves are insensitive to the initial phase~$\xi_0$.}
	\label{fig:lorentz_plane_e0.0_a9_b0_o2_p0_gx_mx}
\end{figure}

Fig.~\ref{fig:lorentz_plane_e0.5_a9_b0_o2_p0_gx_mx} shows an example of \LP with $a=\numprint{e9}$, injection factors $\log \gamma_0=\{0,1,2,3\}$ and varying initial phase~$\xi_0$, $\xi_0=0$ in dashed lines, $\xi_0=\upi/4$ in dotted lines and $\xi_0=\upi/2$ in solid lines. Contrary to a plane wave, injection at high speed reduces the asymptotic Lorentz factor compared to a particle injected at rest. This effect is particularly visible for the \LP and $\xi_0=0$. Indeed at initial high Lorentz factors, the particle does not feel any electromagnetic field because it almost exactly catches up the wave in its node where $\mathbf{B}=\mathbf{E}=\mathbf{0}$. For $\gamma_0=\numprint{e3}$ only after having travelled a distance $\numprint{e3}\,\rlight$ will the particle start to accelerate. If the initial phase differs from zero like for instance $\xi_0=\pi/4$ or $\xi_0=\pi/2$, then the \LP results resemble the \CP evolution shown in Fig.~\ref{fig:lorentz_plane_e0.0_a9_b0_o2_p0_gx_mx} because the particle accelerates right at the injection place $r=\rlight$. For circular polarization, because of the symmetry of the field, the motion remains insensitive to the initial phase of the wave, only the initial Lorentz factor matters.

Simulations including radiation reaction terms in the equation of motion according to the \LL prescription do not alleviate the conclusions drawn above. We indeed checked by inspection of the linear and circular polarization results that the discrepancies are irrelevant.

\subsection{Initial phase of the wave}

The phase at which the particle enters the wave is also arbitrary. Its value $\xi_0$ can vary from injection points to injection points. We already showed some examples in the previous paragraphs. Let us summarize our findings for particles starting at rest at $r_0=\rlight$ and different polarization states.

\begin{figure}
	\centering
	\includegraphics[width=0.95\linewidth]{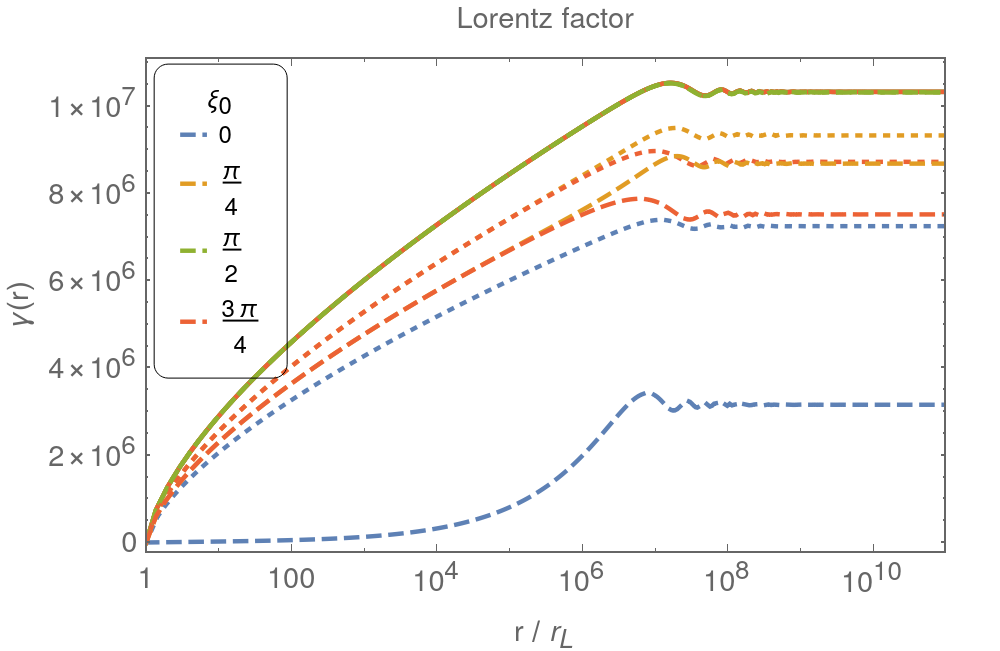}
	\caption{Evolution of the Lorentz factor for circularly, elliptically and linearly polarized waves respectively in solid, dotted and dashed line, with $a=\numprint{e9}$ and several initial phases~$\xi_0$.}
	\label{fig:lorentz_plane_ex_a9_b0_o2_px_g0_mx}
\end{figure}

Fig.~\ref{fig:lorentz_plane_ex_a9_b0_o2_px_g0_mx} shows a particle entering the wave with $a=\numprint{e9}$ and different initial phases $\xi_0=\{0, \upi/4, \upi/2, 3\,\upi/4\}$  for circularly, elliptically and linearly polarized waves respectively in solid, dotted and dashed line. All things being equal, \LP is always the least efficient configuration to energize charged particles. 
If radiation reaction is included, 
we checked that nothing changes significantly again. 

The radiation feedback never perturbs the motion of a charged particle in a spherical wave on the pulsar wind zone. To a very good approximation, this perturbation as implemented in the \LL prescription, is irrelevant in such a case. A last configuration of interest concerns the "collision" between the pulsar large amplitude low frequency vacuum wave with an incoming charged particle. We refer to this process as a head on collision and investigate it in the following closing paragraph.

\subsection{Head on collision}

Particles catching up the spherical wave is less efficient than particles hitting this wave in head on "collision". We therefore also investigated the propagation of particles travelling towards the neutron star, permeating its electromagnetic field. So let us consider a particle coming from infinity. In our runs, it means particles starting at sufficiently large distances where the electromagnetic field has sufficiently decreased to become negligible for the particles to follow straight lines. The particle moves at a relativistic speed with initial Lorentz factor~$\gamma_0$, in the negative $\ex$ direction. Concretely, we also fixed the large distance to $r/\rlight=\numprint{e12}$ at time $t=0$. The particle travels towards the star, feeling an outgoing wave with an increasing strength parameter~$a$. At some distance~$r_{\rm min}$, the electromagnetic field overcomes the particle inertia and turns it back into the positive $\ex$ direction. The minimal distance of approach~$r_{\rm min}$ depends on the initial particle energy.

\begin{figure}
	\centering
	\includegraphics[width=0.95\linewidth]{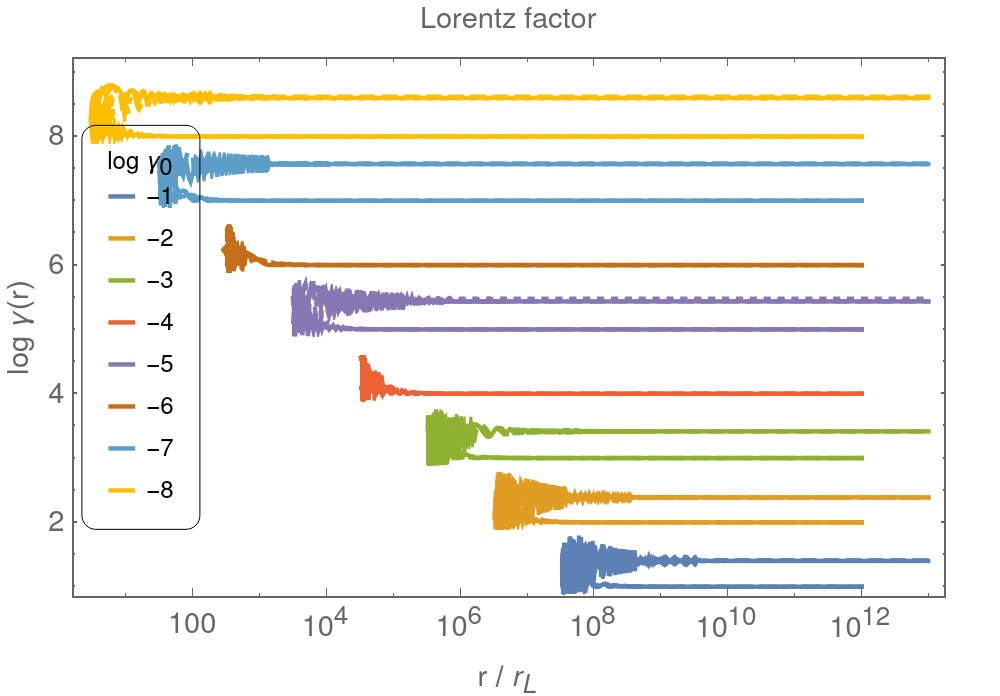}
	\caption{Evolution of the Lorentz factor for a linearly polarized wave with $a=\numprint{e9}$, initial Lorentz factor $\log \gamma_0=i$ with $i\in[1..8]$ and varying initial phases, $\xi_0=0$ in dashed lines, $\xi_0=\upi/4$ in dotted lines and $\xi_0=\upi/2$ in solid lines.}
	\label{fig:lorentz_plane_r12_e0.5_a9_b0_o2_p0_g-x_mx}
\end{figure}

\begin{figure}
	\centering
	\includegraphics[width=0.95\linewidth]{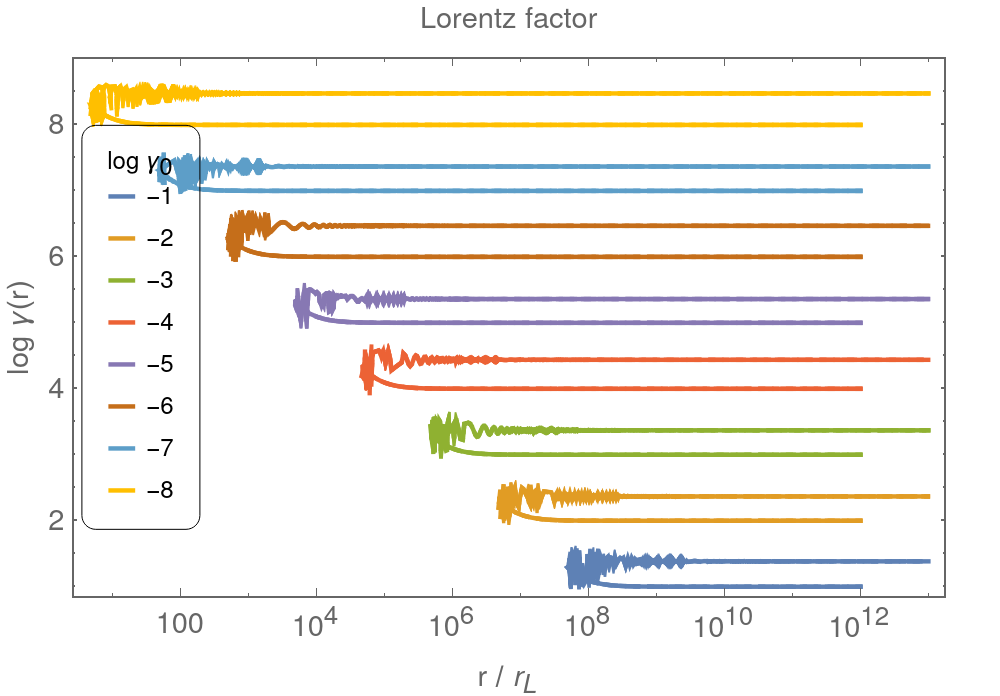}
	\caption{Same as Fig.~\ref{fig:lorentz_plane_r12_e0.5_a9_b0_o2_p0_g-x_mx} but for circular polarization.}
	\label{fig:lorentz_plane_r12_e0.0_a9_b0_o2_p0_g-x_mx}
\end{figure}

As an example, we injected particles with initial Lorentz factors $\log \gamma_0=i$ with $i\in[1..8]$ in a wave of strength $a=\numprint{e9}$. Fig.~\ref{fig:lorentz_plane_r12_e0.5_a9_b0_o2_p0_g-x_mx} shows the evolution of the Lorentz factor for counter-propagating particles and several initial phases~$\xi_0$ for linear polarization. Fig.~\ref{fig:lorentz_plane_r12_e0.0_a9_b0_o2_p0_g-x_mx} shows the equivalent evolution for circular polarization. The gain in energy after bouncing back is irrelevant and independent of the initial phase when entering the wave. It is about a factor 2.5 for all runs.

A simple picture helps to understand the small gain in energy. Let us assume a particle moving in vacuum in the negative $\ex$ direction with Lorentz factor~$\gamma_0$. At $x=0$, it enters a region $x<0$ of constant electromagnetic field with $\mathbf{E}$ directed along the positive $\ey$ direction and $\mathbf{B}$ directed along the positive $\ez$ direction. For such constant fields, exact analytical solutions are known and given for instance by \cite{petri_relativistic_2020}. The particle is deflected by the magnetic part meanwhile accelerated by the electric part. The particle comes out of the electromagnetic field back to the vacuum region $x>0$ with a velocity along the positive $\ex$ direction. It can be shown that the final Lorentz factor after escape is related to the initial speed $\beta_0$ by
\begin{equation}\label{key}
\gamma_{\rm fin} = (1+ 3\,\beta_0) \, \gamma_0 = \gamma_0 + 3 \, \sqrt{\gamma_0^2-1} \approx 4 \, \gamma_0.
\end{equation}
Therefore we find a factor~4 not to different from the factor~2.5 in view of the simple picture we used. This conclusion holds irrespective of the sign of the charge.

\begin{figure}
	\centering
	\includegraphics[width=0.95\linewidth]{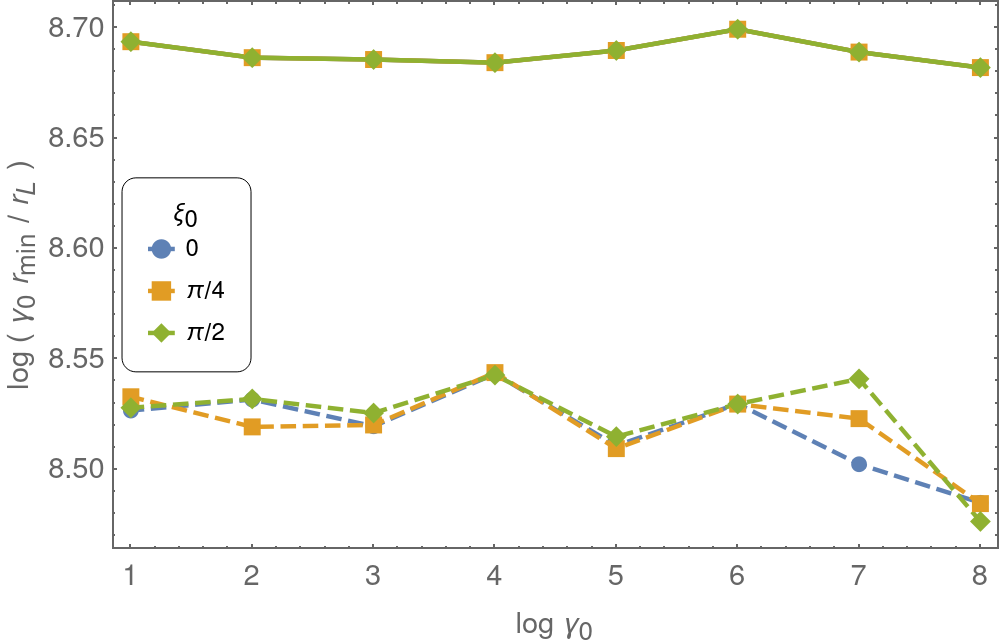}
	\caption{Minimal distance of approach and Lorentz factor for a circularly and linearly polarized wave with $a=\numprint{e9}$ and initial Lorentz factor $\log \gamma_0=i$ with $i\in[1..8]$. The curves are slightly sensitive to the initial phase~$\xi_0$ for \LP.}
	\label{fig:lorentz_plane_rmin_ex_a9_b0_o2_p0_g-x_mx}
\end{figure}

Fig.~\ref{fig:lorentz_plane_rmin_ex_a9_b0_o2_p0_g-x_mx} shows the minimal distance of approach~$r_{\rm min}$ depending on the initial Lorentz factor~$\gamma_0$ and polarization state. A good fit is given by
\begin{equation}\label{key}
 \log \left( \frac{r_{\rm min}}{\rlight}\right) \approx (8.7)_{\rm circ} / (8.5)_{\rm lin} - \log \gamma_0
\end{equation}
the constant value depends on the polarization state, circular or linear.
This minimum distance can be estimated by noting that the particle turns back whenever its Larmor radius $r_{\rm B} = \gamma_0\,c/\omega_{\rm B}$ is comparable to the wavelength of order $\rlight$. In such a situation, the particle performs a half turn in an approximately constant electromagnetic field. Equalling both values leads to
\begin{equation}\label{key}
\gamma_0 \approx \frac{\omega_{\rm B}}{\omega} = a = a_0 \, \frac{\rlight}{r_{\rm min}} .
\end{equation}
In other words, the product $\gamma_0 \, r_{\rm min}$ remains constant and equal to $a_0 \, \rlight$ that is approximately $\numprint{e9}$. The energy gain in this head on collision remains therefore also too weak to account for any acceleration process.

\section{Discussion}
\label{sec:Discussion}

We considered exclusively waves with zero electromagnetic invariants which seems far from reality around a neutron star. However a plane wave solution represents an excellent approximation to the electromagnetic field felt by an ultra-relativistic particle in its rest frame \citep{ritus_quantum_1985}. Therefore the zero electromagnetic invariants assumption is a useful simple case to compute approximate solutions in the ultra-relativistic regime. For instance in low density laser plasma simulations, the field is that of a plane wave, therefore zero invariants apply to high accuracy if the plasma current feedback is neglected. Moreover, if particles move at ultra-relativistic speeds, as in high intensity laser experiments or around neutron stars, in their rest frame the two electromagnetic invariants~$I_1$ and $I_2$ nearly vanish. Indeed, their normalized magnitude defined by
\begin{subequations}\label{key}
\begin{align}
\frac{\mathbf{E}' \cdot \mathbf{B}'}{E'^2+c^2\,B'^2} & = \frac{\mathbf{E} \cdot \mathbf{B}}{E'^2+c^2\,B'^2} \propto \frac{1}{\gamma^2} \ll 1 \\
\frac{E'^2 - c^2 \, B'^2}{E'^2+c^2\,B'^2} & = \frac{E^2 - c^2 \, B^2}{E'^2+c^2\,B'^2} \propto \frac{1}{\gamma^2} \ll 1
\end{align}
\end{subequations}
decrease as $1/\gamma^2$ where $\gamma$ is the particle Lorentz factor in the observer frame. This approximation breaks down only in very special configurations, for instance when particle velocity, electric field and magnetic field are all collinear. This approximation called "locally constant crossed field approximation (LCFA)" is extensively used in the computation of QED effects in laser experiments.

Nevertheless, we emphasize that in a pulsar magnetohydrodynamical (MHD) wind, the electromagnetic invariants are not exactly equal to zero. The solutions given in the previous sections can only barely represent the more realistic situation for a relativistically magnetized outflow. For instance, in ideal MHD where the plasma possesses an infinite conductivity, the electric field vanishes in the plasma rest frame and the wind structure is well approximated by the split monopole solution of \cite{bogovalov_physics_1999}. More generally speaking, particle acceleration in relativistic magnetized outflows is central to the explanation of gamma-ray bursts (GRB). The composition in neutrons and protons and their dynamics impacts on the observational appearance of the GRBs as shown by \cite{derishev_neutron_1999}. Moreover, relativistic jets in blazars can efficiently convert bulk kinetic energy into radiation as found by \cite{stern_radiation_2008}. Similar problems and outcomes are discussed by \cite{beskin_radio_2018} who summarizes the history of pulsar theory development. He points out the importance of the outflow mass load and its particle content made essentially of electrons and positrons with high pair multiplicities to understand the dynamics of the pulsar wind.
\cite{prokofev_primary_2015} showed that in an MHD wind with non vanishing invariants, not only acceleration, but also deceleration of particles is possible. \cite{nokhrina_acceleration_2017} also considered the case of particle injection with arbitrary energies. These ideas are now also supported by PIC simulations seen by \cite{philippov_ab-initio_2018}, \cite{sironi_particle_2017} or \cite{cerutti_dissipation_2020}. Clearly, the variety of solutions is much richer for non null fields and required a deeper investigation.

Our particle injection scheme from an arbitrary point in the wave assumes their birth at that point. A more careful analysis would require an investigation of their entire trajectory, starting from the vicinity of the stellar surface where the electromagnetic field resembles more to a quasi static dipole magnetic field and quadrupole electric field. Physically, particles are injected via electron/positron pair creation through magnetic photon absorption \citep{erber_high-energy_1966} or photon-photon interaction through the Breit-Wheeler process \citep{breit_collision_1934}. This injection mechanism associated with pair production has been widely discussed in the literature. Magnetic photon absorption occurs mainly around the polar caps \citep{sturrock_model_1971, ruderman_theory_1975, alber_cascade_1975, fawley_potential_1977} where the magnetic field is strong enough to disintegrate a high energetic photon. Outer gaps \citep{cheng_energetic_1986} are the privileged sites for the photon-photon interaction although the pulsar striped wind becomes a serious alternative \citep{lyubarskii_model_1996}. \cite{cheng_particle_1980} envisaged even an ion outflow from the polar caps. Traditionally, the acceleration process starts at the birth place and goes on smoothly up to the light-cylinder or further. Nevertheless, in some circumstances, \cite{beskin_particle_2000} found an efficient and abrupt acceleration phase in a narrow band around the light-cylinder. The injection problem is crucial for the outcome of kinetic pulsar magnetosphere simulations. Particles can be extracted right at the surface \citep{wada_particle_2011} or everywhere within the light-cylinder as done by \cite{chen_electrodynamics_2014} for an axisymmetric magnetosphere. Both injection schemes lead to very different stationary states. The role of the particle injection rate was studied by \cite{kalapotharakos_three-dimensional_2018}, see also \cite{brambilla_electronpositron_2018}. On a more fundamental side, \cite{timokhin_maximum_2019} performed a careful analysis of the pair production efficiency, updating their previous work presented in \cite{timokhin_polar_2015}.

The places where particles enter the wave and their associated kinetic energy at injection into this wave determines the large scale motion towards the termination shock. The whole story of particle production, propagation, radiation and mixing into the interstellar medium requires a careful bottom-up analysis encompassing the smallest and the largest time and spatial scales. This preliminary work was only intended to explore the propagation and radiation part in the large amplitude low frequency electromagnetic wave.

\section{Conclusions}
\label{sec:Conclusions}

Neutron stars are believed to be efficient particle accelerators. However, this acceleration process must be quantified depending on the magnetosphere model, being vacuum, force-free or dissipative as well as on radiation feedback. Moreover, realistic physical parameters are required in order to avoid artificial down-scaling of the problem. In this paper we proposed a new approach to tackle those difficult tasks. First we designed an algorithm to solve analytically and semi-analytically for the particle equation of motion in the \LL approximation checking it on known solutions. Next we applied it to spherical waves as those launched by a rotating neutron star. We found that the acceleration efficiency depends on the wave polarization state, strength parameter and on the particle injection conditions, that is its initial speed when entering the wave and the wave initial phase. Because the spherical wave amplitude decreases outside the light-cylinder, we found no evidence of significant radiation damping in the wave zone except in the immediate vicinity of the light-cylinder.

We plan to extend our analysis to waves possessing an electromagnetic field component along the direction of propagation in order to apply it to the exact solution of a magnetic dipole rotating in vacuum and known as Deutsch solution. In such configurations, the light-like electromagnetic field approximation fails and the constant electromagnetic field approximation must be used to treat the most general geometry. The full 3D nature of the problem could then also be incorporated in order to study particle velocities deviating from the wave propagation direction.

Last but not least, the plasma content of the magnetosphere must be taken into account for the most realistic and self-consistent electromagnetic field/particle/radiation interaction. We plan to study test particle motion in those dissipative magnetospheres as found for instance by \cite{petri_electrodynamics_2020}.

\section*{Acknowledgements}

I am grateful to the referee for helpful comments and suggestions. This work has been supported by the CEFIPRA grant IFC/F5904-B/2018 and ANR-20-CE31-0010.

\section*{Data availability}

The data underlying this article will be shared on reasonable request to the corresponding author.





%




%


\bsp	
\label{lastpage}
\end{document}